\documentclass[a4paper,11pt]{article}
\pdfoutput=1
\usepackage{jinstpub}
\usepackage{setspace}
\usepackage{lineno}

\usepackage[normalem]{ulem}
\usepackage{color}
\definecolor{blue}{rgb}{0,0,1}
\definecolor{red}{rgb}{1,0,0}

\newcommand{\e}{\mathrm{e}}
\newcommand{\D}{\mathrm{d}}
\newcommand{\p}{\tilde{p}}
\newcommand{\f}{\tilde{f}}
\newcommand{\F}{\tilde{F}}
\newcommand{\eff}{\mathrm{eff}}

\begin{document}

\title{\boldmath Single electron multiplication distribution in GEM avalanches}

\author[a,1]{Andr\'as L\'aszl\'o,\note{Corresponding author.}}
\author[a,b]{Gerg\H o Hamar,}
\author[a]{G\'abor Kiss}
\author[a]{and Dezs\H o Varga}

\affiliation[a]{Wigner Research Centre for Physics, Budapest, Hungary}
\affiliation[b]{INFN Trieste, Trieste, Italy}

\emailAdd{laszlo.andras@wigner.mta.hu, hamar.gergo@wigner.mta.hu, kiss.gabor@wigner.mta.hu, varga.dezso@wigner.mta.hu}

\abstract{
In this paper, measurement results and experimental methodology are
presented on the determination of multiplication distributions of
avalanches initiated by single electron in GEM foils. The
measurement relies on the amplification of photoelectrons by the GEM
under study, which is subsequently amplified in an MWPC for signal enhancement and 
readout. The intrinsic
detector resolution, namely the sigma-over-mean ratio of the multiplication 
distribution is also elaborated.  Small gain dependence of the shape of the 
avalanche response distribution is observed in the range of net effective gain of $15$
to $100$. The distribution has an exponentially decaying tail at large
amplitudes. At small amplitudes, the applied working gas is seen to have a well
visible effect on the shape of the multiplication distribution. 
Equivalently, the working gas has an influence on the
intrinsic detector resolution of GEMs via suppression of the low
amplitude responses. A sigma-over-mean ratio of $0.75$ was
reached using a neon based mixture, whereas other gases provided an
intrinsic detector resolution closer to $1$, meaning a multiplication
distribution closer to the low-field limit exponential case.
}


\arxivnumber{1605.06939}

\maketitle
\flushbottom

\section{Introduction}
\label{introduction}

The intrinsic fluctuation of electron multiplication in avalanche processes 
of gaseous detectors is an important input parameter at the design phase. An example is the Time Projection Chambers (TPCs) for charged 
particle trajectory detection, where particle identification is intended to 
be performed via specific ionization ($\D{E}/\D{x}$) measurement. The intrinsic energy resolution of the detector, i.e.\ the magnitude of the gain fluctuation, has direct impact on the $\D{E}/\D{x}$ resolution, and hence on the discrimination power between different 
particle masses. For the traditionally used Multi Wire proportional Chamber (MWPC) 
based signal amplification \cite{curran1949}, after fundamental understanding of the relevant processes \cite{raether1964}, avalanche fluctuations have been studied long ago \cite{alkhazov1969, alkhazov1970}, 
revealing a nearly but not precisely exponential multiplication distribution, 
meaning a close to $1$ intrinsic detector resolution in terms of sigma-over-mean 
ratio for single electron response.

Avalanche fluctuations are best measured directly, initiated by a
single electron. This technique, however, is not trivial in the region of low
multiplication, i.e.\ below gains of $\approx 10^3$. In a work of Zerguerras et al
\cite{zerguerras2009, zerguerras2015} such study was carried out for
Micromegas based detectors, revealing a very clear departure from
exponential behavior. In the present paper, we experimentally study
the electron multiplication processes in GEM \cite{sauli2016} based
detector.  In the context of the TPC upgrade project \cite{alicetpc}
of the ALICE experiment at CERN, there is a clear motivation to get
access to such fundamental ingredients of gaseous detectors. The
results can have impact on the ever improving simulations as well
\cite{garfieldpp}. Earlier studies related to GEM gains and intrinsic resolution of 
the GEM multiplication process were also reported in \cite{varga2016}, however 
that experimental setting was not aimed at individual avalanche initiation via a 
single-electron source as in the present paper.

A significant difference of the micropattern based electron multiplication 
relative to the MWPCs is the constrained avalanche evolution space. That potentially results in a deviation from 
the exponential avalanche population distribution, being a limiting case for low field avalanches.
Indeed, for Micromegas detectors a significant deviation from the exponential 
distribution was seen \cite{zerguerras2009, zerguerras2015}, in particular 
a typical sigma-over-mean ratio significantly smaller than $1$ was observed. 
For GEM based amplification processes, a similar phenomenon is potentially 
expected. Our experimental setting was designed in order to quantitatively 
determine the GEM avalanche multiplication distribution and its properties 
in a wide range of effective gains (about $15$ to $100$), and in various working gases, 
with a special emphasis on the low multiplication tail, giving the major 
contribution to the deviation from the exponential distribution. 

The structure of the paper is as follows. 
In Section~\ref{experimentalconfiguration} the experimental configuration is outlined. 
Based on this, in Section~\ref{signalformation} the ingredients of the 
formation of the final recorded detector signal in our setting is detailed. 
In Section~\ref{effective} a brief discussion is dedicated to the coupling effect 
between the GEM inefficiency and the true photoelectron yield, which becomes an 
important detail in the data analysis. 
In Section~\ref{analysis} the data analysis 
methods and consistency checks are detailed. 
In Section~\ref{measurements} the obtained measurement results are reported.
In Section~\ref{conclusion} a summary is provided.

\section{Experimental configuration}
\label{experimentalconfiguration}

\begin{figure}[!ht]
\begin{center}
\includegraphics[width=9cm]{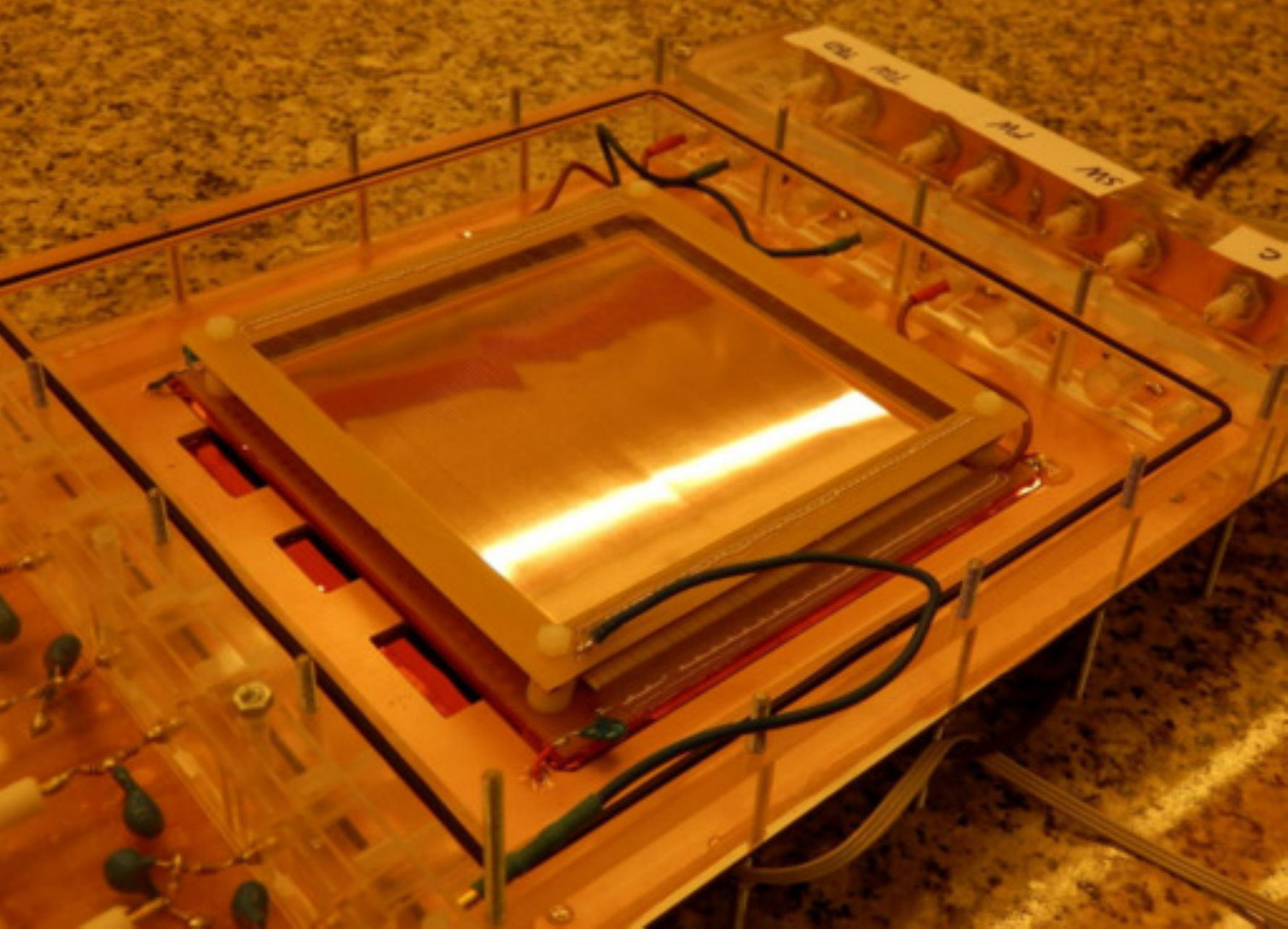}
\end{center}
\caption{(Color online) 
Photograph of the chamber of the experimental setup.}
\label{setupphoto}
\end{figure}

\begin{figure}[!ht]
\begin{center}
\includegraphics[width=9.5cm]{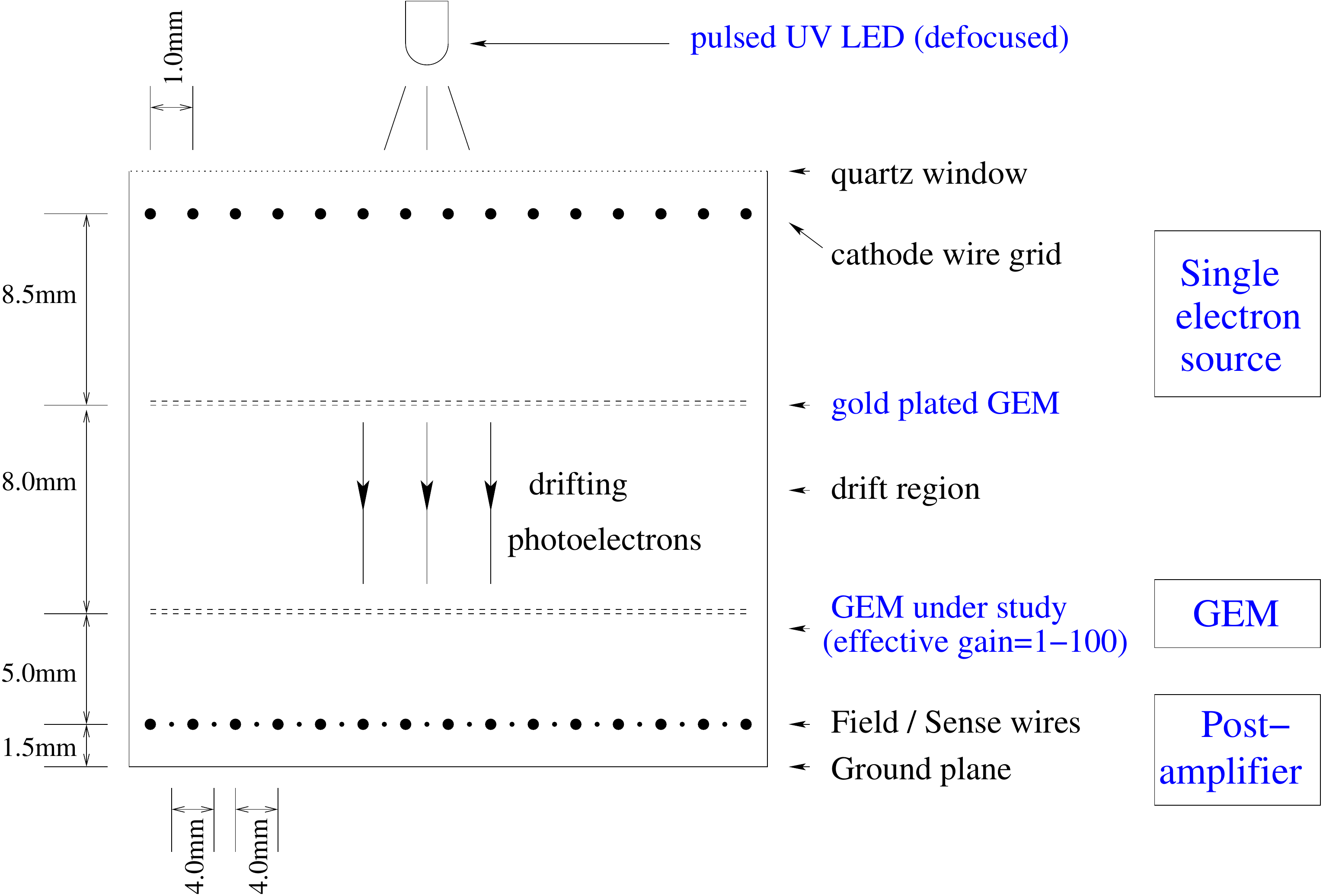}
\end{center}
\caption{(Color online) 
Sketch of the full experimental setup. The most important 
components, namely the pulsed UV LED, the subsequent gold plated GEM as a 
photoelectron source, the studied amplifier GEM, and the high gain 
readout stage of an MWPC region is emphasized on the drawing.}
\label{setup}
\end{figure}

\begin{figure}[!ht]
\begin{center}
\includegraphics{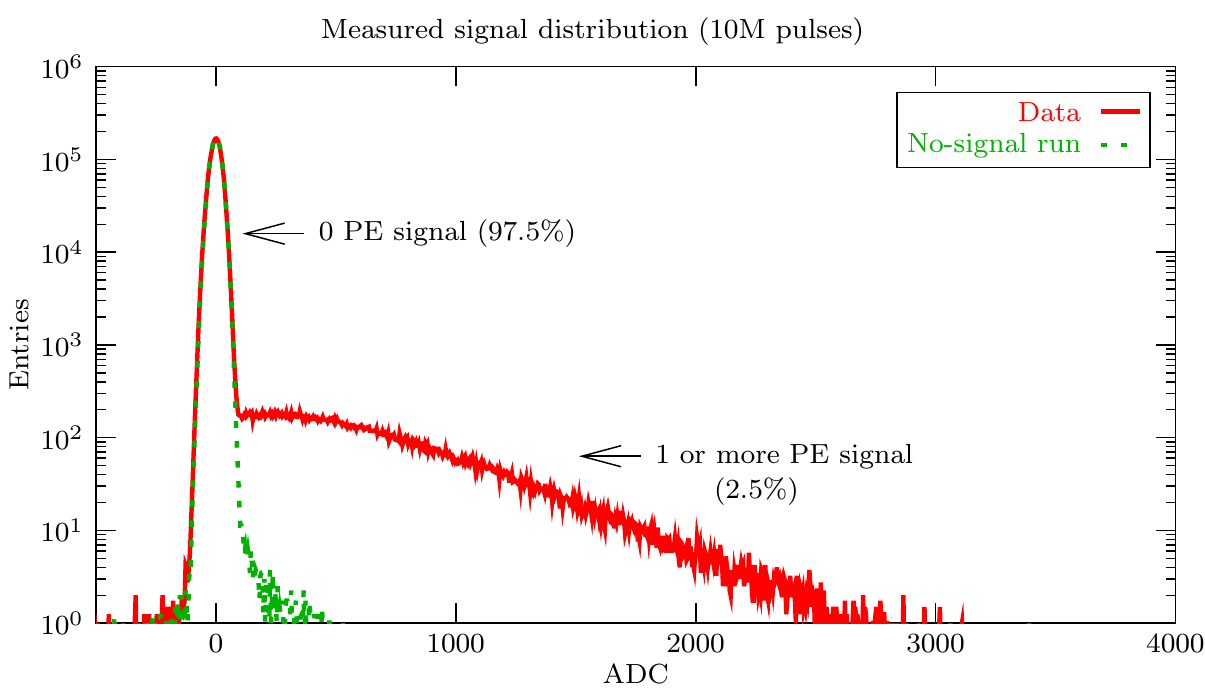}
\end{center}
\caption{(Color online) 
Typical amplitude distribution for UV LED pulses. The peak around 
zero amplitude corresponds to zero-photoelectron events, whereas the 
exponential-like tail at larger amplitudes corresponds to the contribution of 
$1$ or more photoelectron events. In the shown example low photoelectron yield 
was applied, of the order of $2.5\cdot 10^{-2}$ photoelectron per pulse, and therefore 
the relative contribution of multi-photoelectron events to the spectrum is 
negligible. Specifically, the part corresponding to the $1$ or more photoelectron 
response reflects the shape of the $1$ photoelectron response distribution, within a 
systematic error of about a percent. 
The amplitude distribution of a no-signal run is also shown for comparison. 
The corresponding distribution is seen to be a Gaussian with a 
small ``skirt'' --- the latter is due to the cosmic ray background. 
The shown example distribution was recorded in 
Ne(90)CO${}_{2}$(10) working gas, at per pulse photoelectron yield of 
$0.025$, GEM gain of $39$, and MWPC gain of $2346$ in terms of electron 
multiplication.}
\label{rawspectrum}
\end{figure}

The experimental setup is outlined in Fig.~\ref{setupphoto} and \ref{setup}, from which one can
identify the three sections of the system. First is the ``single electron source'', 
second is the actual GEM under study, from which the avalanche is extracted to the
third part, the ``post-amplifier''. 
A typically observed signal distribution is shown in Fig.~\ref{rawspectrum}. 

Specifically, a $22\,\mathrm{kHz}$ 
oscillator trigger was used to pulse a UV LED, which released photoelectrons 
(PEs) from the surface of a gold plated GEM foil inside the working gas volume. 
The electric potential on the gold plated GEM was configured such that the foil 
became transparent (effective gain $\approx 1$) to the released 
PEs, thus providing a source of drifting 
PEs used for the measurements. The mean number of PEs per pulse 
was variable in the range of about $10^{-2}$ to $3$. After the drift region 
the conventional GEM foil under study was placed with an effective gain 
variable in a wide range of about $1$ to $100$-fold multiplication. 
Following the GEM foil under study, a high gain stage was realized by an 
MWPC region, providing further signal enhancement with an electron 
multiplication up to $300$ to $6000$.

The readout Front-End Electronics (FEE) and the LED pulser electronics were originally developed 
for the Leopard project \cite{leopard}. The FEE was triggered using the same 
oscillator trigger as the UV LED, thus providing a zero-bias triggering 
scheme. The corresponding trigger delay between the LED and the FEE 
was adjusted such that the amplified analog signal amplitude was sampled at 
the signal maximum by the FEE. The MWPC design used the Close-Cathode Chamber (CCC) concept 
\cite{ccc1, ccc2} providing a robust amplification region before detection by 
the FEE. 
The FEE reading out the Sense Wires delivered the signal amplitude
in terms of ADCs, with a gain of about 150 electron/ADC.
Typical event statistics 
of $10^{7}$ pulses for each setting were recorded, therefore the statistical errors 
of the measurements are negligible and thus are systematics dominated. 
Various working gases, namely Ar(80)CO${}_{2}$(20), CH${}_{4}$, Ne(90)CO${}_{2}$(10) and 
Ne(90)CO${}_{2}$(10)N${}_{2}$(5) were studied. The pressure of the 
chamber was atmospheric. The typical drift field was $875\,\mathrm{V}/\mathrm{cm}$, 
whereas the transfer field for extracting the GEM-multiplied 
electrons was of the order of $800$ to $900\,\mathrm{V}/\mathrm{cm}$, depending 
on the particular setting. The studied GEM foil was standard double-mask 
etched version with a Kapton\textsuperscript{\textregistered} thickness of $50\,\mu\mathrm{m}$ and copper thickness 
of $5\,\mu\mathrm{m}$ on the two sides. The holes were of the usual 
double-conical shape with inner diameter of $50\,\mu\mathrm{m}$, 
outer diameter of $70\,\mu\mathrm{m}$, and with a pitch of $140\,\mu\mathrm{m}$ in a 
triangular mesh.

Due to the presence of the MWPC ``post-amplifier'' a quite natural question is 
its contribution to the spread of the GEM response distribution. In particular: 
to what extent a raw amplitude spectrum like Fig.~\ref{rawspectrum} reflects 
the shape of the GEM response and to what extent the MWPC response. In 
Section~\ref{signalformation} it is analytically and in Section~\ref{analysis} 
it is quantitatively shown that at larger GEM responses the additional spreading 
effect of the MWPC stage becomes small due to the law of large numbers. In 
addition, it is shown that the spectra can be corrected for the MWPC 
contribution in an exact manner.

During the data recording runs, special care was taken to make sure that the 
parameters of the experimental setup do not drift in time. The time stability 
of the UV LED was checked using photomultiplier measurements. 
The stability of the fields on the GEM was also made sure, so that neither the 
PE yield nor the GEM gain had any drift in time. For this purpose, 
before the physics runs, special chargeup \cite{chargeup} runs were performed with large 
PE yield, of the order of $1$ PE per pulse, and with large GEM gain, of the 
order of $50$, as well as with large MWPC gain, of the order of $3000$. 
The chargeup runs were continued until the amplitude distribution was seen to 
be stabilized as a function of time. 
Additionally, in order to rule out any possibility of a residual drift by the chargeup effects 
during the physics data taking runs, each physics data taking sequence 
was performed two times after each-other in order to explicitly verify 
the time stability of the raw data used in the physics analysis.

\section{Signal formation}
\label{signalformation}

The three well defined stages of the experimental setup allow to identify a 
clear relation between the avalanche distribution and the
measurable signal. This framework is described in the present section.

\subsection{Single electron source}
\label{sigleelectronsource}

Upon a pulse of the UV LED, PEs are emitted from 
the top electrode of the upper, gold plated GEM foil. The number of emitted 
PEs follows a Poisson distribution with an unknown but fixed 
expectation value for a given setting. 
The field on the electrodes of the upper, gold plated GEM 
foil was set in such a way that it became transparent 
(effective gain $\approx 1$), and therefore together with the UV LED the 
upper, gold plated GEM foil acted as a source of PEs, drifting homogeneously 
along the drift field as shown in Fig.~\ref{setup}. The distribution of the 
number $n$ of these PEs are denoted by $P_{\nu}$, being Poissonian with expectation 
value $\nu$, i.e.
\begin{eqnarray}
P_{\nu}(n)=\frac{\e^{-\nu}\nu^{n}}{n!}.
\label{poisson}
\end{eqnarray}

\subsection{Avalanche in the GEM}
\label{avalancheinthegem}

The PEs reaching the amplifier GEM+MWPC region initiate independent 
superimposed responses. Therefore, 
if the amplitude response distribution of the amplifier GEM+MWPC region to a 
single PE were described by a probability distribution $f$ (that is, the probability of 
$x$ electrons being detected by the readout is $f(x)$), then 
together with the fluctuation of the PE statistics, the probability 
distribution of the amplitude response $x$ to a single UV LED pulse would be
\begin{eqnarray}
\sum_{n=0}^{\infty} f^{\star(n)}(x) \,P_{\nu}(n),
\label{totalidealresponse}
\end{eqnarray}
where the symbol $\star$ denotes convolution, and for each non-negative integer 
$n$ the symbol $f^{\star(n)}$ denotes $n$-fold convolution of $f$ with itself. 
In probability theory, a distribution of the form Eq.(\ref{totalidealresponse}) 
is called a \emph{Poisson compound of $f$ with parameter $\nu$} \cite{feller1968, haight1967}. 
The effect of multi-PE contribution can be arbitrarily reduced by making measurements 
with very low PE yields $\nu\approx 10^{-2}$ in which case only the $0$-PE and 
the $1$-PE contribution is relevant. Moreover, in Section~\ref{analysis} we show 
an advanced method in comparison to the traditional low-PE measurement, which 
is able to eliminate the multi-PE contribution from a response distribution 
like Eq.(\ref{totalidealresponse}) in an exact manner, based on Fourier analysis 
of the Poisson compound.

Due to the presence of the usual additive, approximately Gaussian electronic noise at the input of the FEE 
amplifier, the observed amplitude distribution can be written as
\begin{eqnarray}
h & = & g \star \,\sum_{n=0}^{\infty} f^{\star(n)} \,P_{\nu}(n),
\label{totalresponse}
\end{eqnarray}
where $g$ describes the distribution of the additive electronic noise and 
$h$ is the final observed amplitude distribution. 
In practice, experimental 
determination of $g$ was done in ``no-signal'' data samples, with having the LED light covered out using a mask.
The modification effect of the 
electronic noise distribution $g$ on the observed distribution $h$ can be reduced 
by suppressing the additive electronic noise at the FEE input as much as 
possible, and by increasing the total amplification of the system with 
respect to the electronic noise level, i.e.\ with respect to the sigma of 
$g$. Moreover, the remaining small residual 
contribution of the electronic noise may even be fully removed via 
Fourier based deconvolution, as detailed in Section~\ref{analysis}.

\subsection{Signals in the post-amplifier}
\label{signalsinthepostamplifier}

\begin{figure}[!ht]
\begin{center}
\includegraphics{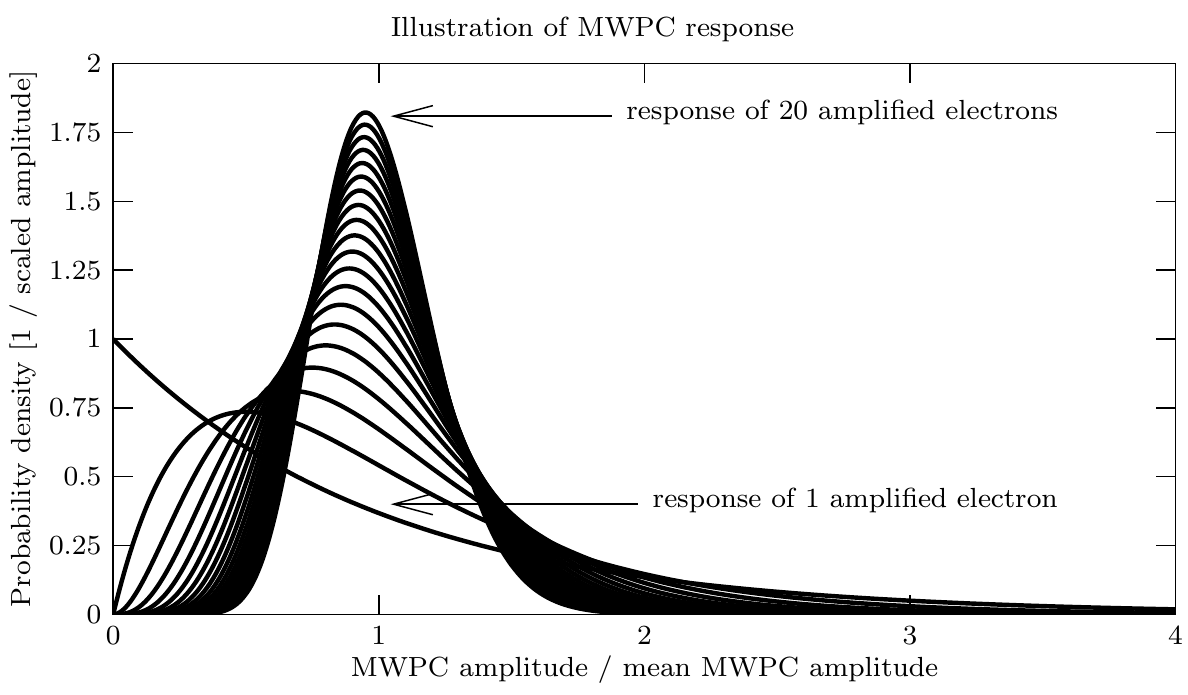}
\end{center}
\caption{(Color online) 
Illustration of the decrease of the relative spread of the MWPC response 
to increasing number of incoming electrons from the GEM region. In particular, 
a realistic model to the MWPC response $e_{\gamma}(x)=1/\gamma\,\exp(-x/\gamma)$ 
to single electron is shown, along with the corresponding response 
$e_{\gamma}^{\star(k)}(x)$ to $k$ electrons. It is seen that the relative width 
of the $k$-electron response goes down with $1/\sqrt{k}$ due to the additivity 
of variance and mean of independent distributions. Because of that, the MWPC 
does not have a large shape distortion effect on the GEM amplitude distribution 
at large GEM responses. In addition, this contribution can be corrected for 
in the spectra in an analytic manner.}
\label{mwpcRespIllustr}
\end{figure}

In Eq.(\ref{totalresponse}) the probability distribution $f$ describes the 
common response of the amplifier GEM+MWPC to a single incoming electron. 
That can be directly related to the distribution of the \emph{effective multiplication response} 
of the studied GEM to a single incoming electron, being the main object 
of interest in the present paper, and denoted by $p$. As such, for a non-negative 
integer $k$, the symbol $p(k)$ denotes the probability of the GEM foil to 
effectively multiply a single incoming electron in such a way that exactly 
$k$ pieces of multiplied electrons can be extracted by the induction field. 
With these notations, the distribution $f$ of the GEM+MWPC response $x$ can be 
written as
\begin{eqnarray}
f(x) & = & \sum_{k=0}^{\infty} e_{\gamma}^{\star(k)}(x) \,p(k),
\label{mwpcplusgemresponse}
\end{eqnarray}
where $e_{\gamma}$ denotes the multiplication response distribution of the MWPC 
with a gain $\gamma$, to a single incoming electron. Similarly as in Eq.(\ref{totalidealresponse}), for any non-negative integer $k$ the 
symbol $e_{\gamma}^{\star(k)}$ denotes $k$-fold convolution of $e_{\gamma}$ with itself. 
The expression $f(x)$ would mean the probability of $x$ electrons to be 
read out from the GEM+MWPC system, given a single incoming electron. Similarly, 
the symbol $e_{\gamma}(x)$ would mean the probability of $x$ electrons to 
be read out from the MWPC, given a single electron entering the MWPC region. 
In consequence, $e_{\gamma}^{\star(k)}(x)$ is the probability of $x$ electrons 
to be read out from the MWPC, given that $k$ electrons entered the MWPC region. 
The model Eq.(\ref{mwpcplusgemresponse}) is suggested by the idea that the MWPC avalanches 
triggered by the extracted GEM-multiplied electrons evolve independently. This 
is justified by the applied relatively low MWPC gains (about $\gamma\approx 300$ to $3500$ in 
terms of electron multiplication) and by the relatively low number of MWPC-amplified 
incoming electrons (GEM gains of the order of $1$ to $100$). In the light of 
Eq.(\ref{mwpcplusgemresponse}) it is an interesting question to ask that given 
the GEM response distribution, how much additional spread is introduced by 
the response of our MWPC post-amplifier? In practical terms: given a raw signal 
spectrum such as in Fig.~\ref{rawspectrum}, to what extent that reflects the 
shape of the GEM response distribution of interest and to what extent there can be some 
shape modification by the MWPC amplification? 
That contribution is illustrated in Fig.~\ref{mwpcRespIllustr}. It is seen 
that the spread of the MWPC response function has less and less influence on 
the amplified GEM spectra at larger GEM amplitudes, since the relative spread 
of the $k$-electron MWPC response goes down as $1/\sqrt{k}$. That is simply 
due to the additivity of variance and mean of independent distributions. 
In Section~\ref{analysis} it is shown that this small effect can be also 
corrected for in an exact manner.

The Leopard FEE \cite{leopard} used for the signal readout from the Sense Wires 
included a preamplifier, having a total gain of $1$ ADC for about $150$ electrons. 
Since the studies performed in the paper needed a relatively large dynamical 
range in terms of measured amplitudes, a careful study was performed to check 
and compensate any possible nonlinearities of the FEE preamplifier when 
approaching the saturation range. This is also detailed in Section~\ref{analysis}.

In the list below the possible contributions of all the known detector effects 
are summarized which can turn up as ingredients in the observed amplitude distribution 
for single UV LED pulses.
\begin{enumerate}
 \item Photoelectron statistics fluctuations $P_{\nu}$.
 \item GEM effective multiplication fluctuations $p$, being the main object of interest of the study presented in the paper.
 \item MWPC multiplication fluctuations $e_{\gamma}$.
 \item Additive electronic noise fluctuations $g$ at the input of the FEE.
 \item A possible deterministic non-linearity transfer function of the FEE amplifier, due to the wide dynamic range of the study.
 \item Pedestal shift and subsequent AD conversion.
\end{enumerate}
In Section~\ref{analysis} the methodologies for quantification and correction 
for the above effects are detailed, in order to obtain measurements for the 
GEM effective multiplication distribution $p$ with the lowest possible systematic 
errors.

\section{Coupling effect between detection inefficiency and photoelectron yield}
\label{effective}

In this section a brief probability theory argument is presented, showing 
that the true PE yield and the GEM inefficiency cannot be disentangled, 
i.e.\ it is not possible to know a priori whether a close-to-zero amplitude was detected because 
no PE was emitted by a given LED pulse, or some PEs were present in the system, 
but the corresponding GEM response was lost due to an intrinsic inefficiency effect. 
The argument shows that the system with possible GEM inefficiencies can, however, 
be uniquely characterized by an effective PE yield, being the product of the 
GEM efficiency and of the true PE yield. 
An inefficiency effect, i.e.\ a high value for $f(0)$, can appear for instance 
when the PE is lost due to attachment in the gas \cite{blum2008}, which 
can be caused by electronegative gas contamination.

Let us introduce the definition
\begin{eqnarray}
\p(k) & = & \frac{1}{1-p(0)}\,\left(1-\delta_{0\,k}\right)\,p(k)
\label{ptilde}
\end{eqnarray}
for all non-negative integers $k$, where $\delta_{0\,k}$ denotes the Kronecker delta 
of index $0,k$. Since $p(0)$ is the probability of extracting $0$ electrons 
from the amplifier GEM foil, i.e.\ the \emph{GEM inefficiency}, $\p$ is 
the conditional probability distribution of the number of extracted 
GEM-multiplied electrons with the condition of extracting at least $1$ electron. 
This quantity, i.e.\ the distribution of the \emph{net effective multiplication}, 
excludes the GEM inefficiency $p(0)$. Using such notation, one has the identity
\begin{eqnarray}
p(k) & = & p(0)\,\delta_{0\,k} + \left(1-p(0)\right)\,\p(k)
\label{psplitting}
\end{eqnarray}
for all non-negative integers $k$. The splitting Eq.(\ref{psplitting}) may also 
be realized at the level of amplification by the combined GEM+MWPC stage, namely with the 
definition
\begin{eqnarray}
\f(x) & = & \sum_{k=0}^{\infty} e_{\gamma}^{\star(k)}(x)\,\p(k)
\label{ftilde}
\end{eqnarray}
one has the identity
\begin{eqnarray}
f(x) & = & p(0)\,\delta(x) + \left(1-p(0)\right)\,\f(x),
\label{fsplitting}
\end{eqnarray}
where $\delta$ denotes the Dirac delta, and $x$ is the amplitude response. It is seen that 
$\f$ is the conditional multiplication distribution of the GEM+MWPC system 
with the condition that at least $1$ electron was extracted from the GEM foil. 
That is, $\f$ is almost the same as $f$, except that the GEM inefficiency 
$p(0)$ is not counted within. Note that $\f$ can be re-expressed as
\begin{eqnarray}
\f(x) & = & \frac{1}{1-p(0)} \,\sum_{k=1}^{\infty} e_{\gamma}^{\star(k)}(x) \,p(k).
\label{ftilde2}
\end{eqnarray}
From such an alternative representation it is seen that $\f$ can be considered as a regular probability density 
function of the continuous variable $x$, since the singular Dirac delta contribution within 
$f$ is excluded from $\f$, and because the MWPC multiplication response 
distribution function $e_{\gamma}$ behaves like a regular probability density 
function of a continuous variable above the applied MWPC gains larger than $\gamma\approx 300$. 
As a consequence of Eq.(\ref{totalresponse}), 
Eq.(\ref{poisson}) and the convolution theorem \cite{arfken1985, bracewell1999}, 
in Fourier space one has the identity
\begin{eqnarray}
H & = & G\,\sum_{n=0}^{\infty} F^{n}\, \frac{\e^{-\nu}\nu^{n}}{n!} \cr
  & = & G\, \exp\left(-\nu\,\left(1-F\right)\right) \cr
  & = & G\, \exp\left(-\left(1-p(0)\right)\nu\,\left(1-\F\right)\right),
\label{Heq}
\end{eqnarray}
where $F$, $\F$, $G$, $H$ are the Fourier transforms of the probability density 
functions $f$, $\f$, $g$, $h$, respectively. Motivated by Eq.(\ref{Heq}), 
the \emph{effective photoelectron yield}
\begin{eqnarray}
\nu_{\eff} & = & \left(1-p(0)\right)\,\nu,
\label{nueff}
\end{eqnarray}
is introduced, which is the true PE yield $\nu$ multiplied by the GEM efficiency $1-p(0)$. 
With that, the identity
\begin{eqnarray}
H & = & G\, \exp\left(-\nu_{\eff}\,\left(1-\F\right)\right)
\label{Heq2}
\end{eqnarray}
follows, which after an inverse Fourier transformation yields an alternative 
form of Eq.(\ref{totalresponse}):
\begin{eqnarray}
h & = & g \star \,\sum_{n=0}^{\infty} \f^{\star(n)} \,P_{\nu_{\eff}}(n).
\label{totalresponse2}
\end{eqnarray}
This identity explicitly shows that in our setting the GEM inefficiency and 
the PE yield cannot be disentangled, but a configuration can still be characterized by the 
effective PE yield $\nu_{\eff}$, and by the net effective multiplication 
distribution $\p$ and $\f$ of the GEM and the GEM+MWPC system. Our analysis 
will thus aim to characterize the net effective GEM multiplication distribution 
$\p$.

\section{Analysis, calibration and data consistency}
\label{analysis}

In each given setting, the distribution of the amplitude response to 
single UV LED pulses in terms of ADCs were recorded. The processes involved 
in the formation of the ADC signal are described in Section~\ref{signalformation} and 
Section~\ref{effective}. 
In this section the analysis procedures, used for correction or quantification 
of those detector effects are detailed step-by-step.

\subsection{Pedestal shift and electronic noise quantification}

The zero point of the AD converter of the FEE is determined via special 
no-signal pedestal runs, with the light from the UV LED blocked. 
Using the pedestal value, the observed amplitude distributions were corrected such that 
zero signal corresponds to zero ADC.  
The pedestal runs also provide measurements on the electronic noise fluctuations, i.e.\ on the 
probability density function $g$ appearing in Eq.(\ref{totalresponse}) 
and Eq.(\ref{totalresponse2}). Typically, $g$ turned out to be an approximately 
Gaussian distribution with a standard deviation of the order of $20\,\mathrm{ADC}$s.

\subsection{FEE non-linearity characterization}

\begin{figure}[!ht]
\begin{center}
\includegraphics{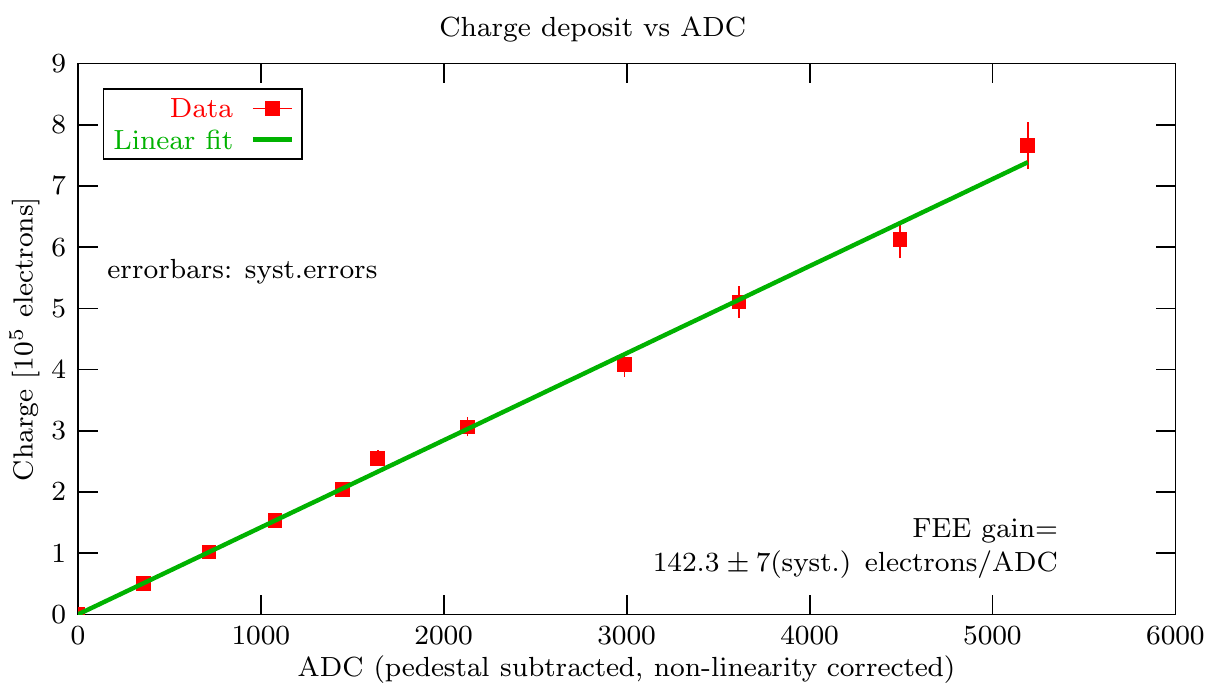}
\end{center}
\caption{(Color online) Deposited charge versus the corresponding ADC. 
The measurement was performed using a test pulser delivering a well defined amount of 
charge on the Sense Wires. 
The shown ADC values are already pedestal subtracted and non-linearity corrected. 
Good linearity is seen in the entire dynamical range. 
The deposited test charge values 
carry approximately $5\%$ systematic errors. This pulser test also provides 
the value of the FEE gain in terms of electron count to ADC conversion, being 
$142.3\pm 7(\mathrm{syst.})\,\mathrm{electrons/ADC}$.}
\label{pulser}
\end{figure}

For accurate determination of the $1$-PE response distribution $\f$ (or $\p$), 
a large dynamical range is needed in order to simultaneously resolve the 
exponential-like tail at large amplitudes and a possible deviation from 
exponential at low amplitudes. Motivated by this, the linearity of the FEE 
was carefully checked.
One possibility is using different MWPC gain settings (that is, at the same GEM avalanche distribution). A residual 
systematic deviation of no more than $0.5\%$ from linearity was observed after the applied 
non-linearity compensation. The linearity after the compensation 
was also validated using direct charge deposition on the Sense Wires with a pulse generator.
The result is shown in Fig.~\ref{pulser}, demonstrating good linearity. 
This test has systematic errors not worse than $5\%$ in relative accuracy due to the uncertainty of the test pulse charge. The measurement readily provides the FEE gain in terms of charge over ADC. 
The measured FEE gain was found to be $142.3\pm 7(\mathrm{syst.})\,\mathrm{electrons/ADC}$.

\subsection{Elimination of multi-PE contribution}

Due to the identity Eq.(\ref{totalresponse2}), the observed amplitude distributions 
have contributions from $0$-PE events with a weight of $\e^{-\nu_{\eff}}=O(\nu_{\eff}{}^{0})$, 
from $1$-PE events with a weight of $\e^{-\nu_{\eff}}\,\nu_{\eff}=O\left(\nu_{\eff}{}^{1}\right)$, and 
multi-PE events with a total weight of $1-\e^{-\nu_{\eff}}\,\left(1+\nu_{\eff}\right)=O\left(\nu_{\eff}{}^{2}\right)$. 
The contamination by the multi-PE events within the sample of $1$ or more PE 
events has thus a dependency of $\frac{1-\e^{-\nu_{\eff}}\,\left(1+\nu_{\eff}\right)}{\e^{-\nu_{\eff}}\,\nu_{\eff}}=O(\nu_{\eff}{}^{1})$. 
Therefore, the most commonly applied method, as also done e.g.\ in \cite{zerguerras2009, zerguerras2015}, 
is to take measurements with very low effective PE yield $\nu_{\eff}\leq 10^{-2}$, 
in which case the multi-PE component of the amplitude distribution $h$ becomes dominated by the 
$1$-PE yield within a systematic error of $0.5\%$. 
The simplicity of this approach makes that method 
a good experimental reference, however, as e.g.\ seen in 
Fig.~\ref{rawspectrum} the separation of the $0$-PE and $1$-PE 
response distributions is not a trivial task due to their substantial overlap.

\begin{figure}[!ht]
\begin{center}
\includegraphics{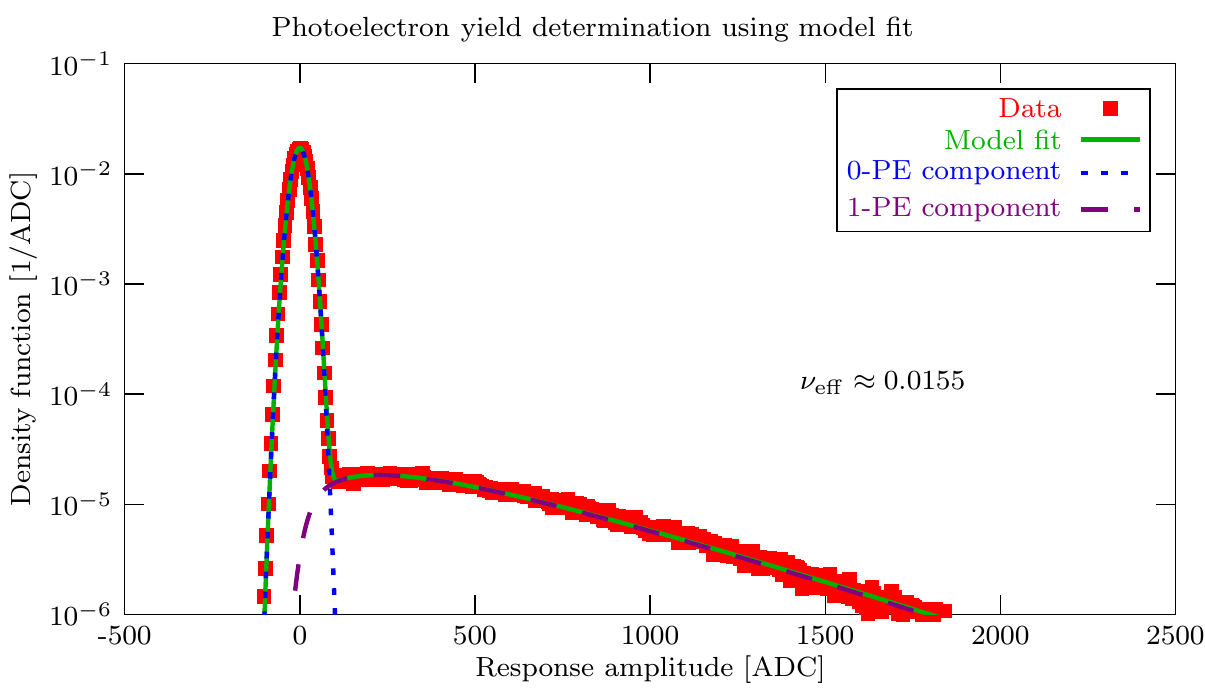}
\end{center}
\caption{(Color online) Disentangling of the $0$-PE and the $1$-PE contribution 
to the amplitude distribution via model fit. A parametric Gamma distribution 
shape is assumed for the true $1$-PE response distribution $\f$. The model 
describes the measured distribution very well. The fit procedure also quantifies 
the effective PE yield $\nu_{\eff}$. The shown example was recorded in 
Ne(90)CO${}_{2}$(10) working gas, at per pulse effective PE yield of 
$0.0155$, GEM gain of $39$, and MWPC gain of $2346$ in terms of electron 
multiplication.}
\label{GaussGammaPoissonFit}
\end{figure}

A possible way to disentangle the $0$-PE and the $1$-PE contribution is to 
fit a model of the form Eq.(\ref{totalresponse2}) to the observed amplitude 
distribution $h$, with some parametric model for the $1$-PE 
distribution $\f$, such as a Gamma distribution of the form 
$1/\left(s\,\Gamma\left(\kappa\right)\right)\,\left(x/s\right)^{\kappa-1}\,\e^{-x/s}$ 
for the response amplitudes $x$. The Gamma distribution is the continuous analogy 
of P\'olya distribution, often used in parametrization of avalanche responses. Here $\Gamma$ denotes the Gamma function, 
$s$ is the slope parameter and the parameter $\kappa$ measures the deviation 
from exponential, namely the sigma-over-mean ratio of such Gamma distribution 
is $1/\sqrt{\kappa}$. This kind of model fit based analysis is shown in 
Fig.~\ref{GaussGammaPoissonFit}. The data is seen to be very well described 
by the Gamma distribution for the $1$-PE response distribution $\f$. The fit also leads to 
an accurate estimate for the effective PE yield $\nu_{\eff}$.

\begin{figure}[!ht]
\begin{center}
\includegraphics{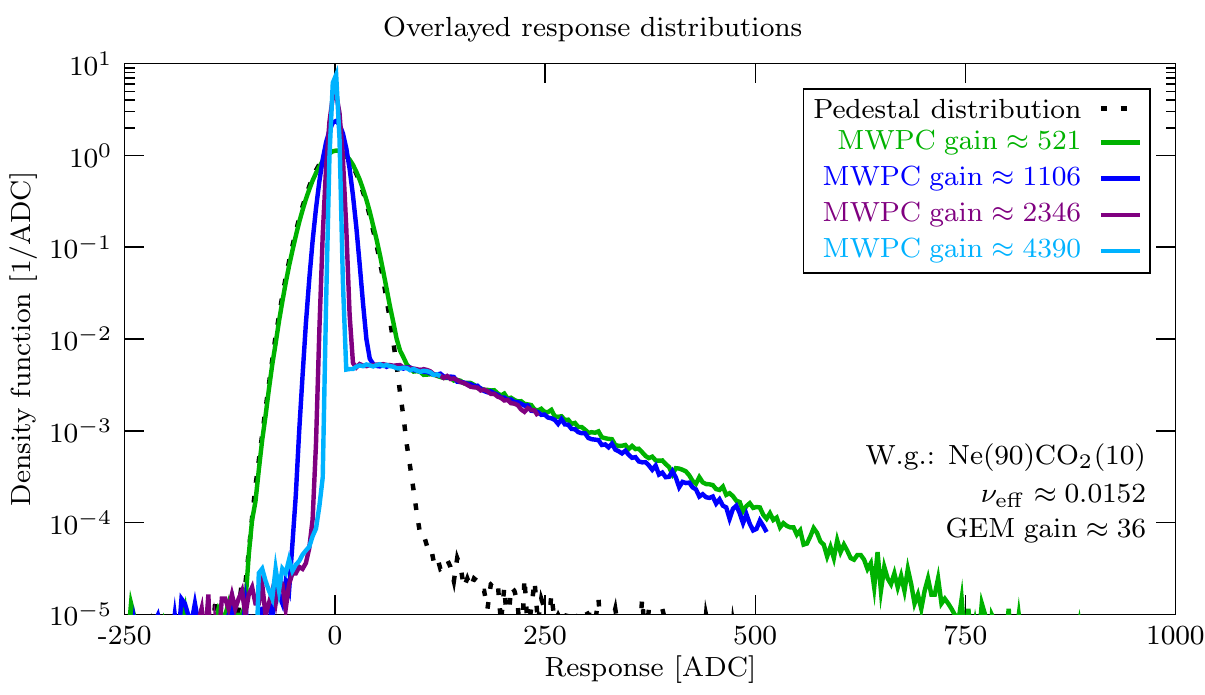}
\end{center}
\caption{(Color online) Disentangling of the $0$-PE and the $1$-PE contribution 
via MWPC gain scan and subsequent overlaying of amplitude distributions after 
rescaling with the relative MWPC gains of the settings. Due to the scan in 
the gain of the final amplifier stage, the MWPC, the low amplitude region 
can be better disentangled from the $0$-PE peak, determined by the electronic noise 
distribution. The shown example was recorded in 
Ne(90)CO${}_{2}$(10) working gas, at per pulse effective PE yield of 
$0.0152$, GEM gain of $36$, and MWPC gain of $521$ to $4390$ in terms of electron 
multiplication.}
\label{MWPCgainscan}
\end{figure}

The model fit method, although it helps to reduce the systematic error in the 
PE yield estimation, brings in a slight model dependence, especially at the 
low amplitude region of the distribution. This can be reduced by recording 
the response distribution $h$ in several copies with all the settings fixed 
except for the MWPC gain. In that case, the overlap of the low 
amplitude region of the $1$-PE response with the electronic noise peak of 
the $0$-PE response can be significantly reduced. This procedure is possible 
due to the presence of the MWPC ``post-amplifier''. 
Such a study is shown in Fig.~\ref{MWPCgainscan}, which helps to experimentally 
get closer to the low amplitude region of the $1$-PE response distribution.

\begin{figure}[!ht]
\begin{center}
\includegraphics{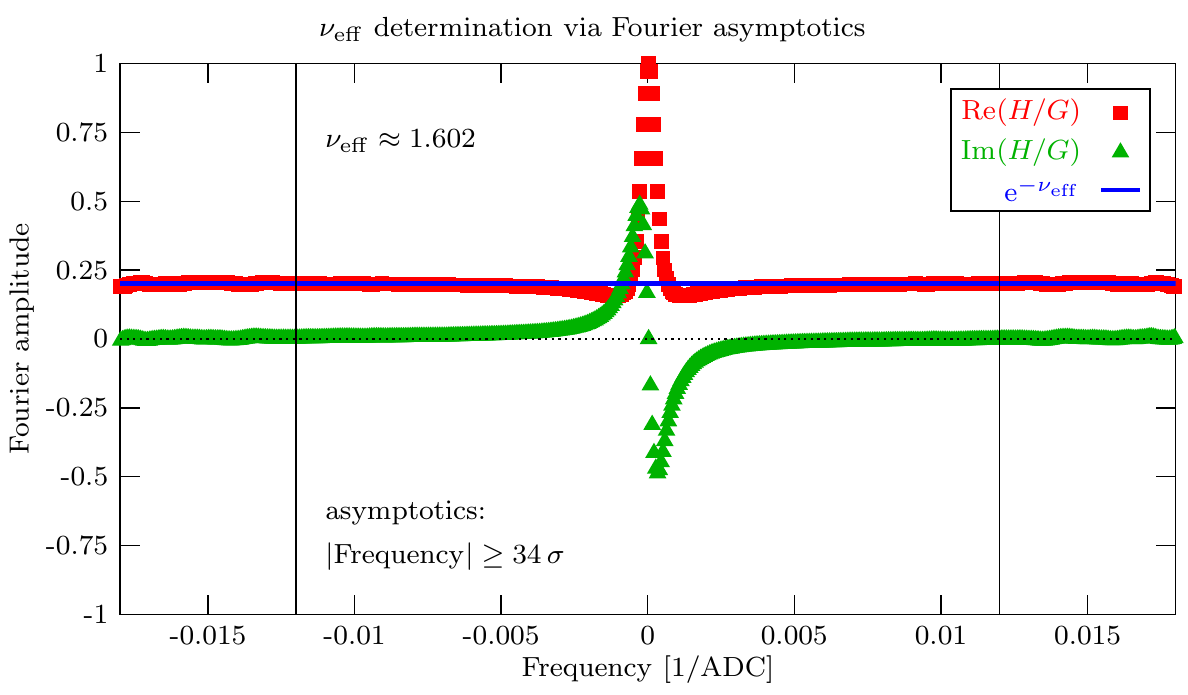}
\end{center}
\caption{(Color online) Experimental determination of the effective PE yield 
$\nu_{\eff}$ using Fourier based Poisson compound decomposition method. The asymptotic value of the Fourier spectrum 
ratio $H/G$ of the amplitude and noise distributions relaxes to $\e^{-\nu_{\eff}}$ 
by means of Eq.(\ref{nudetermination}). The asymptotic region for fitting 
the constant $\e^{-\nu_{\eff}}$ model to the data was a sideband of 
$\left\vert\mathrm{frequency}\right\vert\geq 34\,\sigma$, shown by the vertical lines. 
The shown example was recorded in 
Ne(90)CO${}_{2}$(10) working gas, at per pulse effective PE yield of 
$1.602$, GEM gain of $36$, and MWPC gain of $1106$ in terms of electron 
multiplication.}
\label{numeasurement}
\end{figure}

\begin{figure}[!ht]
\begin{center}
\includegraphics{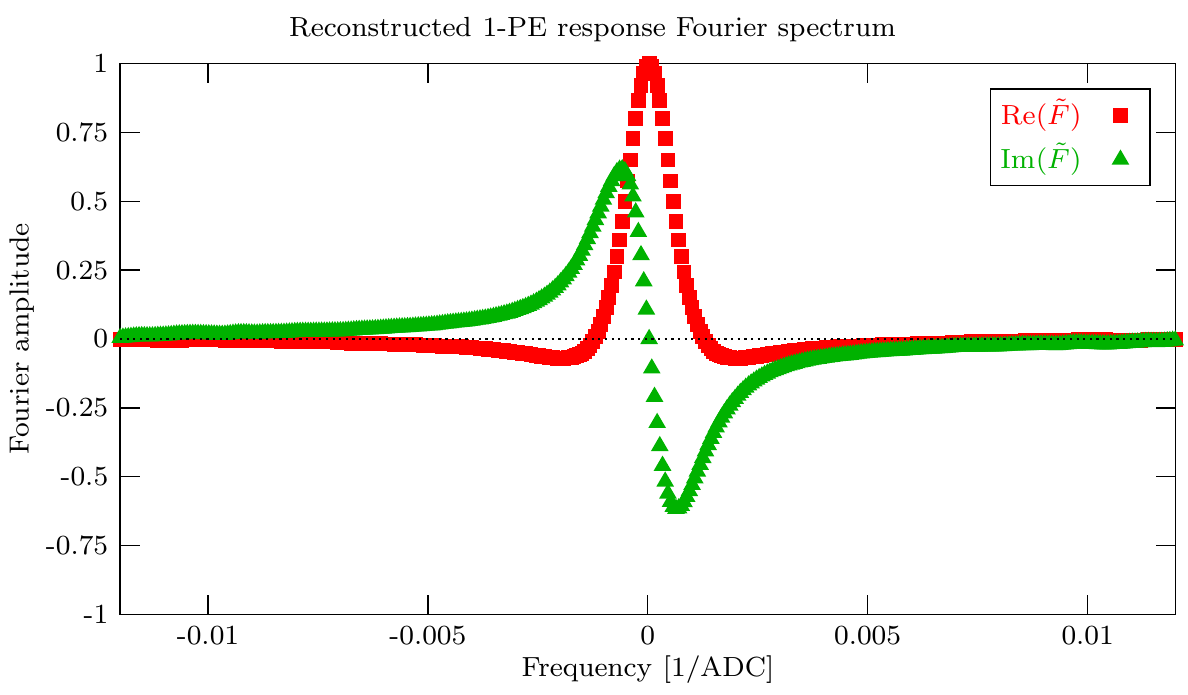}
\end{center}
\caption{(Color online) Experimental determination of the Fourier spectrum 
$\F$ of the pure $1$-PE response distribution $\f$ via Eq.(\ref{Feq}). 
The shown example was recorded in 
Ne(90)CO${}_{2}$(10) working gas, at per pulse effective PE yield of 
$1.602$, GEM gain of $36$, and MWPC gain of $1106$ in terms of electron 
multiplication.}
\label{Fmeasurement}
\end{figure}

\begin{figure}[!ht]
\begin{center}
\includegraphics{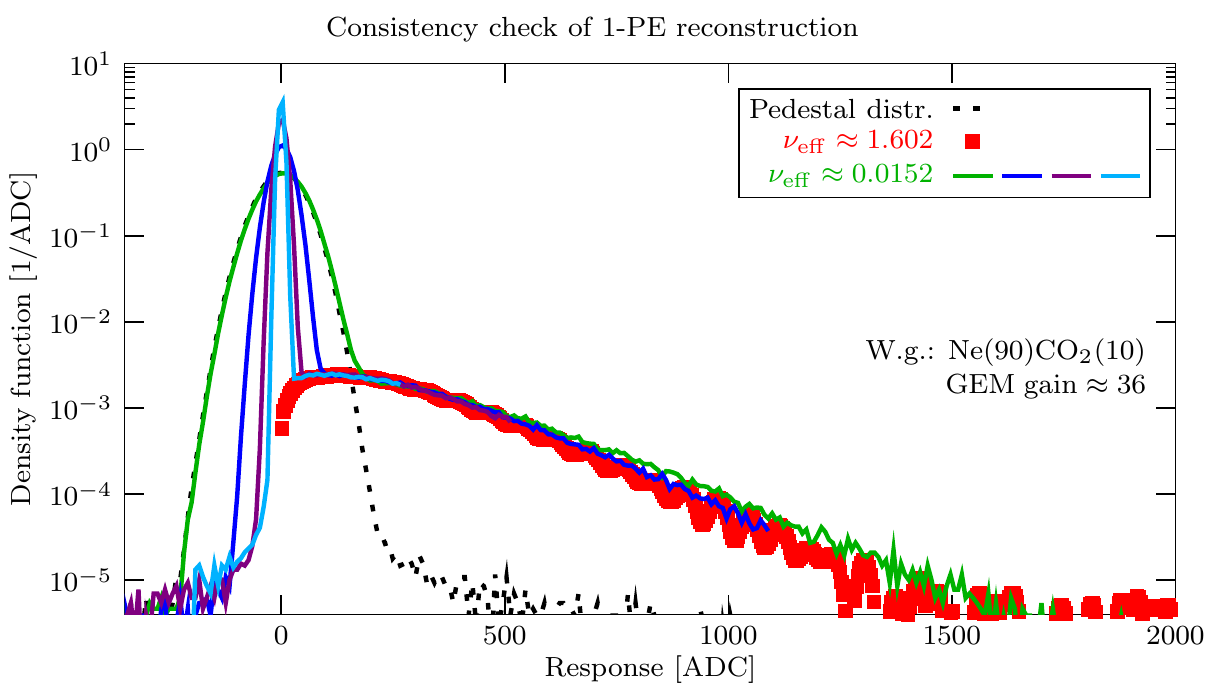}
\end{center}
\caption{(Color online) Experimental determination of the 
pure $1$-PE response distribution $\f$ via Poisson compound decomposition 
method in Fourier space. For cross-check purposes, the more direct measurement 
using MWPC gain scan is also overlayed on the figure, and good agreement is seen. 
The shown example was recorded in 
Ne(90)CO${}_{2}$(10) working gas, at per pulse effective PE yield of 
$1.602$, GEM gain of $36$, and MWPC gain of $1106$ in terms of electron 
multiplication.}
\label{fmeasurement}
\end{figure}

A model independent study of the low amplitude region of the $1$-PE response 
$\f$ at simultaneously fixed effective PE yield, GEM gain and MWPC gain 
can be performed using Eq.(\ref{Heq2}). 
That identity allows one to reconstruct the Fourier spectrum 
$\F$ of the $1$-PE response distribution $\f$, given the Fourier spectrum 
$H$ of the total observed amplitude distribution $h$, the Fourier spectrum 
$G$ of the electronic noise distribution $g$, and the effective PE yield 
$\nu_{\eff}$ via the formula:
\begin{eqnarray}
\F & = & 1+\frac{1}{\nu_{\eff}}\ln\left(\frac{H}{G}\right).
\label{Feq}
\end{eqnarray}
After an inverse Fourier transformation, this delivers the $1$-PE distribution 
of interest $\f$ without model assumptions. Since $h$, $H$, $g$, $G$ can be 
determined from experimental data the only unknown for the procedure to work 
is the effective PE yield $\nu_{\eff}$. Given that, the multi-PE contribution 
can be factorized in an exact manner via Eq.(\ref{Feq}) 
at any PE yield, in particular also at the large PE yield limit. Clearly, 
an increased effective PE yield $\nu_{\eff}$ improves the signal to noise 
ratio, thus the task remains to determine $\nu_{\eff}$ from the 
amplitude distribution $h$ at large PE yields in a reliable way. That can 
be done via the Riemann-Lebesgue lemma \cite{rudin1987}, which states that 
the Fourier spectrum of every probability density function of a continuous 
variable, in particular $\F$, decays to zero at infinite frequencies. 
As a consequence of that and of Eq.(\ref{Heq2}), one has the identity
\begin{eqnarray}
\lim_{\vert\omega\vert\,\rightarrow\,\infty}\, \frac{H(\omega)}{G(\omega)} & = & \e^{-\nu_{\eff}}
\label{nudetermination}
\end{eqnarray}
which can be used for determination of the effective PE yield $\nu_{\eff}$ 
from the asymptotics of the ratio of the Fourier spectrum $H$ of the amplitude 
distribution $h$ and of the Fourier spectrum $G$ of the electronic noise 
distribution $g$. The identity Eq.(\ref{nudetermination}) is illustrated in 
Fig.~\ref{numeasurement}. It is seen that the ratio $H/G$ of the Fourier 
spectra relaxes to a constant value, $\e^{-\nu_{\eff}}$, at the asymptotics, 
thus determining the effective PE yield $\nu_{\eff}$. The asymptotic region 
used for the determination of $\e^{-\nu_{\eff}}$ via fit was defined to be a 
$\left\vert\mathrm{frequency}\right\vert\geq 34\,\sigma$ sideband. Here 
$\sigma=1/(2\pi\,(\sigma_{h}^{2}-\sigma_{g}^{2})^{1/2})$ 
is a lower estimate of the standard deviation of the non-noise component 
of the Fourier spectrum, in which $\sigma_{h}$ and $\sigma_{g}$ denotes the 
standard deviation of the distributions $h$ and $g$, respectively. 
After the determination of the value of $\nu_{\eff}$, the Fourier 
spectrum $\F$ of the pure $1$-PE response $\f$ can be determined via the 
formula Eq.(\ref{Feq}), which is shown in Fig.~\ref{Fmeasurement}. 
The reconstructed pure $1$-PE response distribution 
$\f$ can then be obtained via an inverse Fourier transformation. For 
the Fourier transformations, the FFT implementation of GSL \cite{gsl} was used. 
The resulting reconstructed $1$-PE response distribution is shown in 
Fig.~\ref{fmeasurement}. For cross-check purposes it is overlayed with the 
direct measurement using the MWPC gain scan method shown previously in Fig.~\ref{MWPCgainscan}. 
Good consistency is seen between the two independent approach. The advantage 
of the Fourier based Poisson compound decomposition method is the complete elimination of the overlap region 
with the electronic noise of the $0$-PE contribution and thus the extrapolation 
uncertainty to small amplitudes is not present anymore.

A further possibility of the elimination of the multi-PE contribution in a model independent way 
is performing the analysis at the level of moments. In the low field limit, the avalanche process is self-similar, 
i.e.\ the conditional probability density above any multiplication threshold 
is the same as the full multiplication distribution, which implies that it 
is exponential. An exponential distribution has an important property: 
its sigma-over-mean ratio is $1$. A distribution deviating 
from the exponential, e.g.\  a Gamma distribution, has a sigma-over-mean 
slightly different than that of $1$. Therefore, the sigma-over-mean ratio 
of the $1$-PE distribution provides important information about a deviation 
from the limiting exponential case. Due to the Poisson compound nature 
Eq.(\ref{totalresponse2}) of the observed amplitude distribution $h$, one has the identity 
\begin{eqnarray}
 \mu_{h}        & = & \mu_{g} + \nu_{\eff}\,\mu_{\f},\cr
 \sigma_{h}^{2} & = & \sigma_{g}^{2} + \nu_{\eff}\,\left(\sigma_{\f}^{2}+\mu_{\f}^{2}\right),
\label{mom}
\end{eqnarray}
where $\mu_{h}$, $\sigma_{h}$, $\mu_{g}$, $\sigma_{g}$, $\mu_{\f}$, $\sigma_{\f}$ 
denote the mean and standard deviation of the distributions $h$, $g$ and $\f$, respectively. 
That leads to the equation
\begin{eqnarray}
\frac{\sigma_{\f}}{\mu_{\f}} & = & \sqrt{\nu_{\eff}\,\frac{\sigma_{h}^{2}-\sigma_{g}^{2}}{(\mu_{h}-\mu_{g})^{2}}-1}
\label{sigmapermeanf}
\end{eqnarray}
for the sigma-over-mean ratio of the $1$-PE response distribution $\f$. It is seen 
that solely by determining the effective PE yield $\nu_{\eff}$, the sigma-over-mean 
ratio of the $1$-PE amplitude response distribution $\f$ can simply be 
reconstructed just from the first two statistical moment of the amplitude distribution 
$h$ and of $g$. The results of such statistical moment based analysis shall 
also be presented in Section~\ref{measurements}.

\subsection{Elimination of the MWPC response contribution}

The true motivation of the present study is to measure the GEM response distribution, for which reason
an unfolding procedure has been developed to eliminate the MWPC contribution from
the measured GEM+MWPC signal. This correction is seen to be small, and can 
be performed both at the level of response distributions or at the level of 
moments, i.e.\ specifically for the sigma-over-mean ratios.

The $1$-PE amplitude response distribution of the GEM and the GEM+MWPC system 
is related via Eq.(\ref{ftilde}). It shall be shown that the shape modification 
effect by the MWPC stage is suppressed with increased GEM gain, and is rather 
small above gains of $\approx 10$. This is because for large $k$, 
the distribution $e_{\gamma}^{\star(k)}$ in Eq.(\ref{ftilde}) approximates 
a narrow Gaussian with mean $k\,\gamma$ and standard deviation 
$\sqrt{k}\,\sigma_{e_{\gamma}}$, i.e.\ following the shape of the 
weight $\p$ accurately for large $k$ (see also Fig.~\ref{mwpcRespIllustr}). Moreover, the shape modification 
effect of the MWPC can be corrected in an exact manner, as shall be shown in the following.

\begin{figure}[!ht]
\begin{center}
\includegraphics{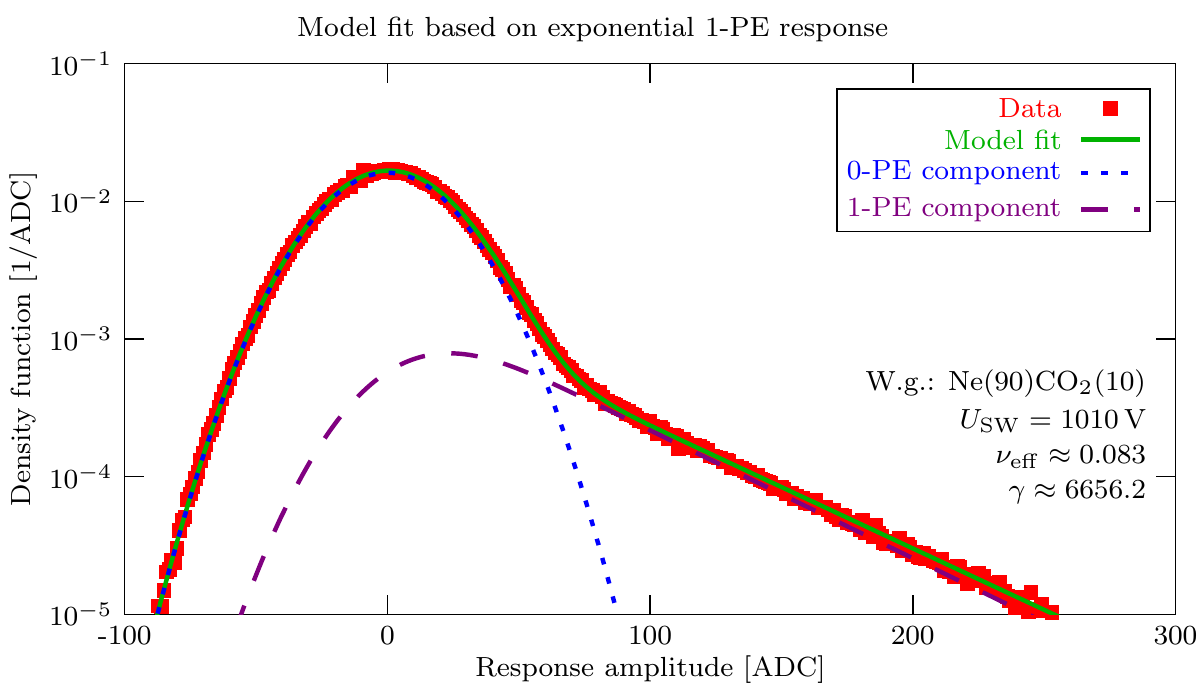}
\end{center}
\caption{(Color online) Model fit to MWPC-only amplitude distribution 
in a typical setting. The used model assumed of the form Eq.(\ref{totalresponse2}) 
with the $1$-PE response distribution $\f$ taken here to be of purely exponential type, 
and the noise distribution $g$ being a Gaussian distribution measured from 
pedestal runs. Apparently, the exponential model for the $1$-PE distribution 
describes the data well. The shown data set was recorded in Ne(90)CO${}_{2}$(10) working 
gas, at net effective PE yield $\nu_{\eff}\approx 0.083$ and at $\gamma\approx6656.2$ 
MWPC gain in terms of electron multiplication, corresponding to 
$U_{\mathrm{SW}}=1010\,\mathrm{V}$ potential on the Sense Wires.}
\label{mwpconlyfit}
\end{figure}

In order to quantify the effect of shape modification of the GEM response by the MWPC stage, one 
should note that the $1$-electron amplitude response distribution $e_{\gamma}$ 
of the MWPC is of exponential type to a good accuracy \cite{alkhazov1969, alkhazov1970} 
at gains of the order of $10^{2}$ to $10^{4}$. This is quite expected due to the 
large number of avalanche generations in the MWPC multiplication process, 
which tends to shift the avalanche evolution closer towards the limiting case. 
The close to exponential nature of MWPC response distributions is 
also confirmed by our control measurements. In such control runs we set 
the amplifier GEM foil to transparent (effective GEM gain $\approx 1$), 
recorded the amplitude distribution with large PE yield, of the order of 
$1$ to $3$ PE per pulse in order to compensate the missing GEM gain in the 
signal to noise ratio. Then, with the method of moments described previously, 
the sigma-over-mean ratio for the $1$-electron response was 
estimated via the formula Eq.(\ref{sigmapermeanf}). That estimate 
resulted in $0.922\pm 0.05(\mathrm{syst.})$ at typical settings, being rather close to the 
exponential having sigma-over-mean ratio of $1$. 
Moreover, the fit of convolution of the Gaussian noise model with the 
Poisson compound of an exponential $1$-PE response distribution of the MWPC 
system describes the MWPC-only data very well, as shown in Fig.~\ref{mwpconlyfit} 
for a typical setting. Thus, an approximation 
of $e_{\gamma}$ with exponential distribution having slope parameter $\gamma$ 
is well justified.

\begin{figure}[!ht]
\begin{center}
\includegraphics{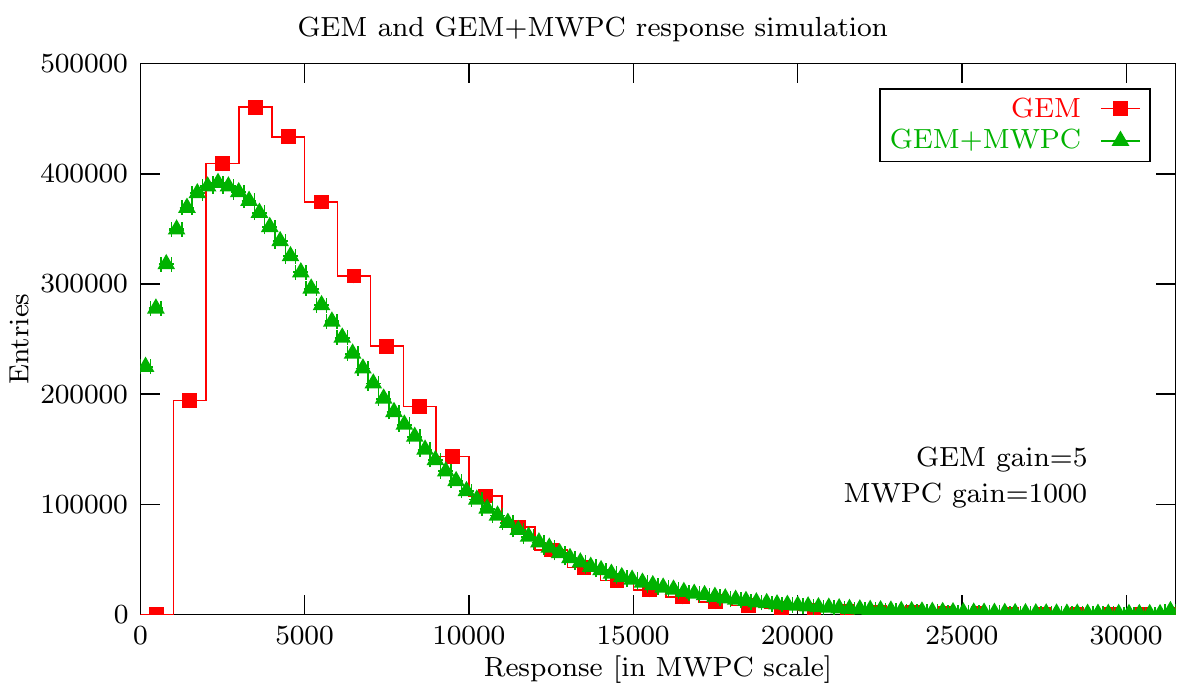}
\includegraphics{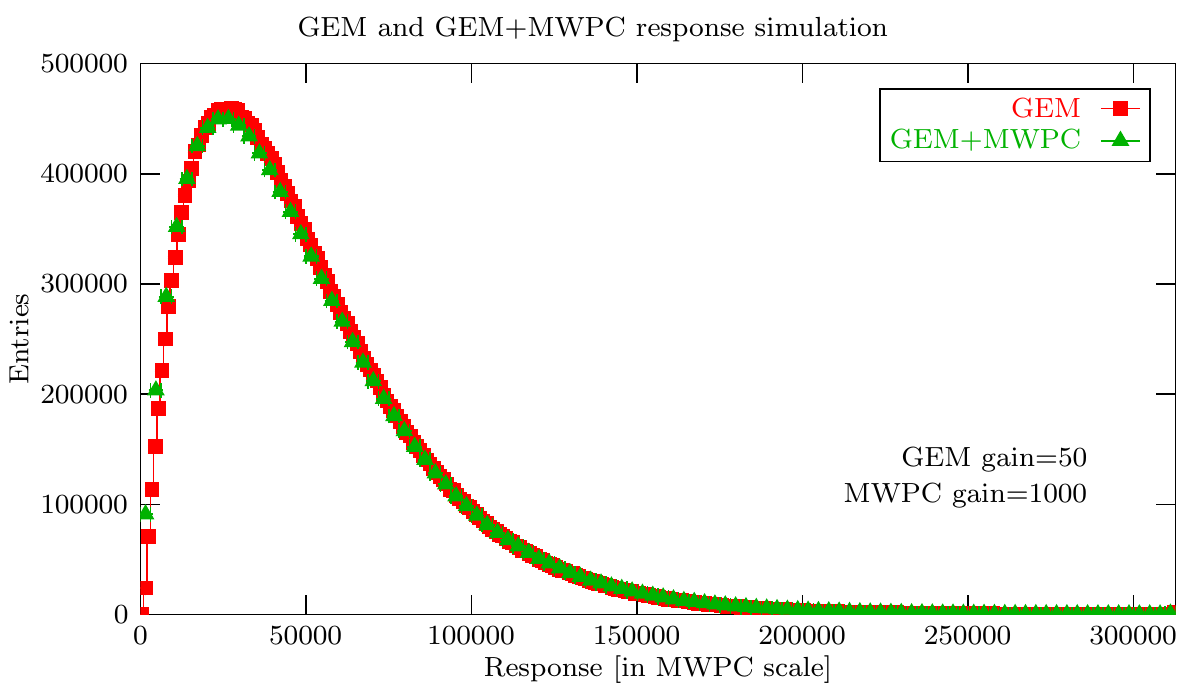}
\end{center}
\caption{(Color online) Top panel: a simulation example in order to demonstrate 
the shape distortion effect of the MWPC response at net GEM effective gain $=5$. 
Bottom panel: the same with net GEM effective gain $=50$. It is seen that 
the shape distortion is negligible at the higher GEM gain setting. The hypothetical 
GEM amplitude distribution was assumed to be a Gamma distribution in both cases, 
with $\kappa=2$.}
\label{mwpcsim}
\end{figure}

The effect of the shape modification by the MWPC stage in terms of Eq.(\ref{ftilde}) 
is illustrated by a simulation shown in Fig.~\ref{mwpcsim}, assuming the $1$-PE 
GEM response to be some Gamma distribution, and $e_{\gamma}$ to be of exponential type. 
The top panel shows that the effect may not be negligible for small 
effective GEM gains, whereas in the bottom panel it is seen that the shape 
distortion effect by the MWPC stage for larger GEM gain settings is negligible.

\begin{figure}[!ht]
\begin{center}
\includegraphics{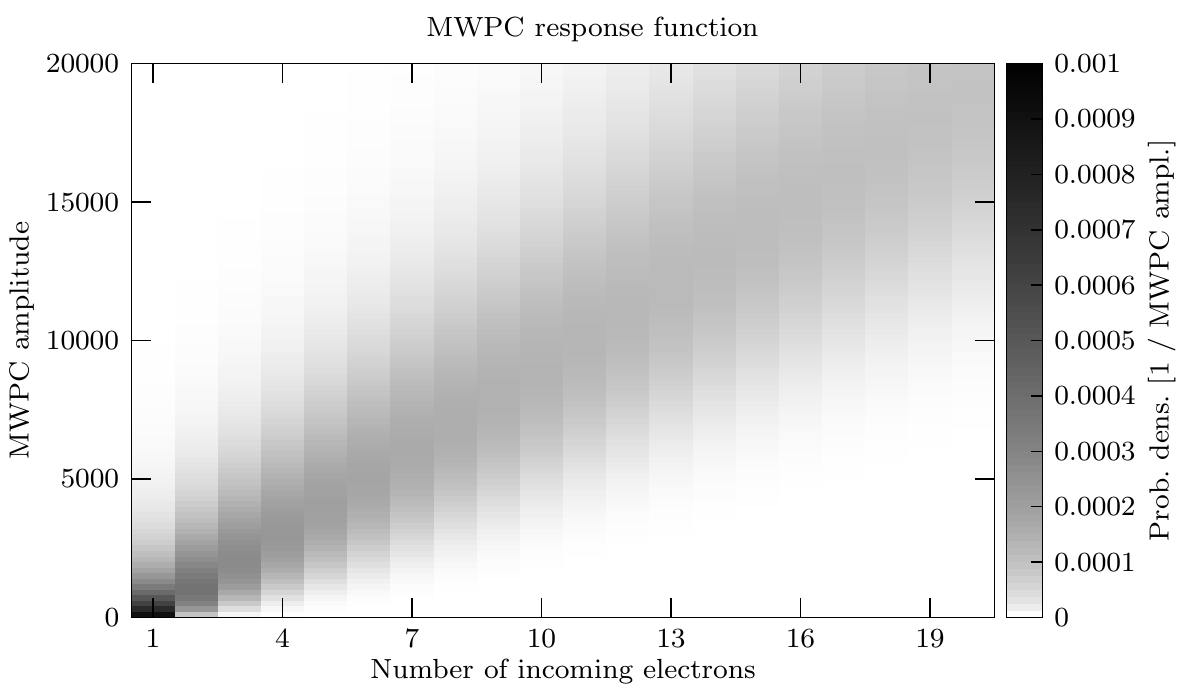}
\includegraphics{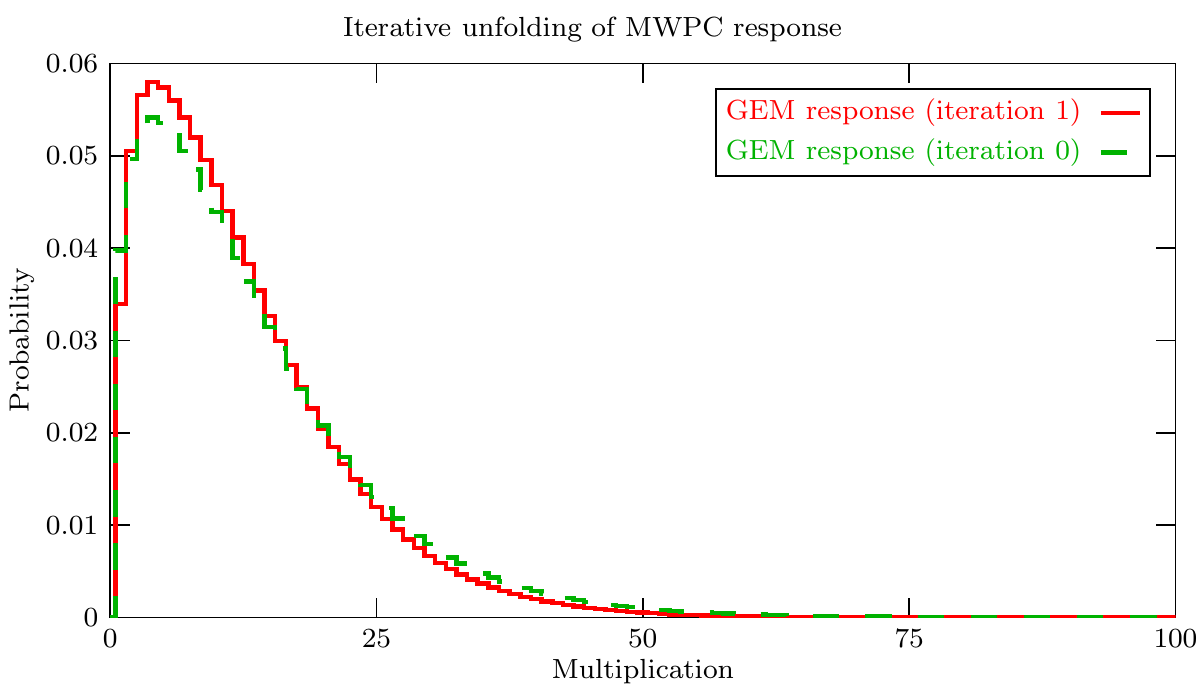}
\end{center}
\caption{(Color online) Top panel: the response function of the 
MWPC amplification stage. Bottom panel: the result of the iterative 
unfolding method \cite{laszlo2016, laszlo2015, laszlo2012} for the removal of 
the MWPC response function from the amplitude distribution. The method 
already converges in one iteration, and the corresponding correction is 
seen to be very small. The shown data was recorded in 
Ne(90)CO${}_{2}$(10) working gas, at GEM gain $12$, and MWPC gain $3015$ 
in terms of electron multiplication.}
\label{respfunc}
\end{figure}

Whenever the shape distortion effect by the MWPC is not considered to be fully 
negligible, it may be corrected for in an exact manner using unfolding. 
That is because Eq.(\ref{ftilde}) reflects that $\f$ is nothing but $\p$ 
folded with a response function of the form 
$\rho_{\gamma}(x \vert k)=e_{\gamma}^{\star(k)}(x)$ for non-negative integers 
$k$ and amplitude values $x$. That kind of transformations can be inverted 
using unfolding methods such as \cite{laszlo2016, laszlo2015, laszlo2012}. 
The pertinent response function is illustrated in Fig.~\ref{respfunc} as well as 
the result of the iterative unfolding by the method \cite{laszlo2016, laszlo2015, laszlo2012}.

The effect of the MWPC response can also be removed at the level of moments. 
Namely, because of Eq.(\ref{ftilde}), the identity
\begin{eqnarray}
\mu_{\f}        & = & \mu_{e_{\gamma}}\,\mu_{\p}, \cr
\sigma_{\f}^{2} & = & \sigma_{e_{\gamma}}^{2}\,\mu_{\p} + \mu_{e_{\gamma}}^{2}\,\sigma_{\p}^{2}
\end{eqnarray}
follows, where $\mu_{\f}$, $\sigma_{\f}$, $\mu_{e_{\gamma}}$, $\sigma_{e_{\gamma}}$, $\mu_{\p}$, $\sigma_{\p}$ 
denote the mean and standard deviation of the distributions 
$\f$, $e_{\gamma}$ and $\p$, respectively. Taking into account that by 
definition one has $\mu_{e_{\gamma}}=\gamma$, the expression
\begin{eqnarray}
\frac{\sigma_{\p}}{\mu_{\p}} & = & \sqrt{\left(\frac{\sigma_{\f}}{\mu_{\f}}\right)^{2}-\left(\frac{\sigma_{e_{\gamma}}}{\mu_{e_{\gamma}}}\right)^{2}\,\frac{\gamma}{\mu_{\f}}}
\label{sigmapermeanp}
\end{eqnarray}
follows for the sigma-over-mean ratio of the $1$-electron net GEM 
multiplication distribution $\p$. From the right hand side it is seen that the 
correction for the contribution of the MWPC response is suppressed by $1$ 
/ GEM gain, i.e.\ by $1/\mu_{\p} = \gamma/\mu_{\f}$. This quantitatively 
demonstrates the phenomenon that the MWPC response does not significantly 
distort the shape of the GEM distribution at GEM gains larger than about 
$10$. Putting together the formula Eq.(\ref{sigmapermeanf}) and 
Eq.(\ref{sigmapermeanp}), one arrives at
\begin{eqnarray}
\frac{\sigma_{\p}}{\mu_{\p}} & = & \sqrt{\nu_{\eff}\,\frac{\sigma_{h}^{2}-\sigma_{g}^{2}}{(\mu_{h}-\mu_{g})^{2}}-\left(\frac{\sigma_{e_{\gamma}}}{\mu_{e_{\gamma}}}\right)^{2}\,\frac{\nu_{\eff}\,\gamma}{\mu_{h}-\mu_{g}}-1},\cr
 & & 
\label{sigmapermeanpfinal}
\end{eqnarray}
which can further be simplified using $\sigma_{e_{\gamma}}/\mu_{e_{\gamma}}\approx 1$. 
In summary: the sigma-over-mean ratio of the $1$-electron GEM response can 
be experimentally determined via measuring $\mu_{h}$, $\sigma_{h}$, $\mu_{g}$, $\sigma_{g}$, 
$\nu_{\eff}$ and $\gamma$ in a given setting. The result of such moment 
based analysis shall also be shown in Section~\ref{measurements}.

\subsection{Estimation of GEM and MWPC gains}

For the interpretation of the obtained results, estimation of GEM gain $\mu_{\p}$ 
in a given setting is necessary. Also, as shown previously, for the corrections 
for the MWPC effects having an estimate for the MWPC gain $\gamma$ is also needed. 
For an estimation procedure, the identity
\begin{eqnarray}
\mu_{h}-\mu_{g} & = & \nu_{\eff}\,\gamma\,\mu_{\p}
\label{gain}
\end{eqnarray}
is the starting point, which means that by measuring $\mu_{h}$, $\mu_{g}$ and 
$\nu_{\eff}$, the combined GEM+MWPC gain $\gamma\,\mu_{\p}$ is readily available. 
Our procedure was to obtain calibration curve of the MWPC gain $\gamma$ as a 
function of Sense Wire voltage in order to calculate $\mu_{\p}$ from the combined 
gain $\gamma\,\mu_{\p}$ in a given setting.

\begin{figure}[!ht]
\begin{center}
\includegraphics{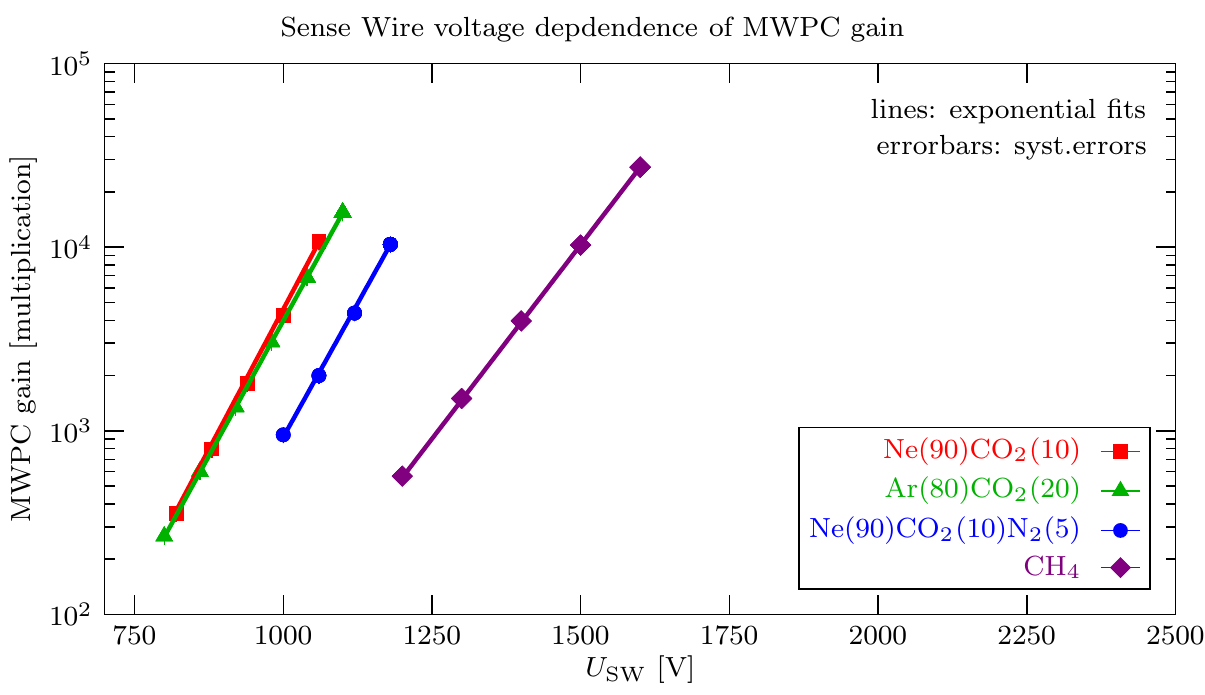}
\end{center}
\caption{(Color online) The determined MWPC gain $\gamma$ curves as a function of 
Sense Wire voltage $U_{\mathrm{SW}}$, in various working gases. The shape of 
the curves were determined at a fixed large effective PE yield ($\approx 1$) 
and at fixed large GEM gain ($\approx 50$), using Sense Wire voltage scan. The normalization of the curves 
were done at large effective PE yield ($\approx 1$) and with transparent 
GEM setting (net effective gain $\approx 1$), via Eq.(\ref{mwpcgainnorm}).}
\label{mwpcgains}
\end{figure}

\begin{figure}[!ht]
\begin{center}
\includegraphics{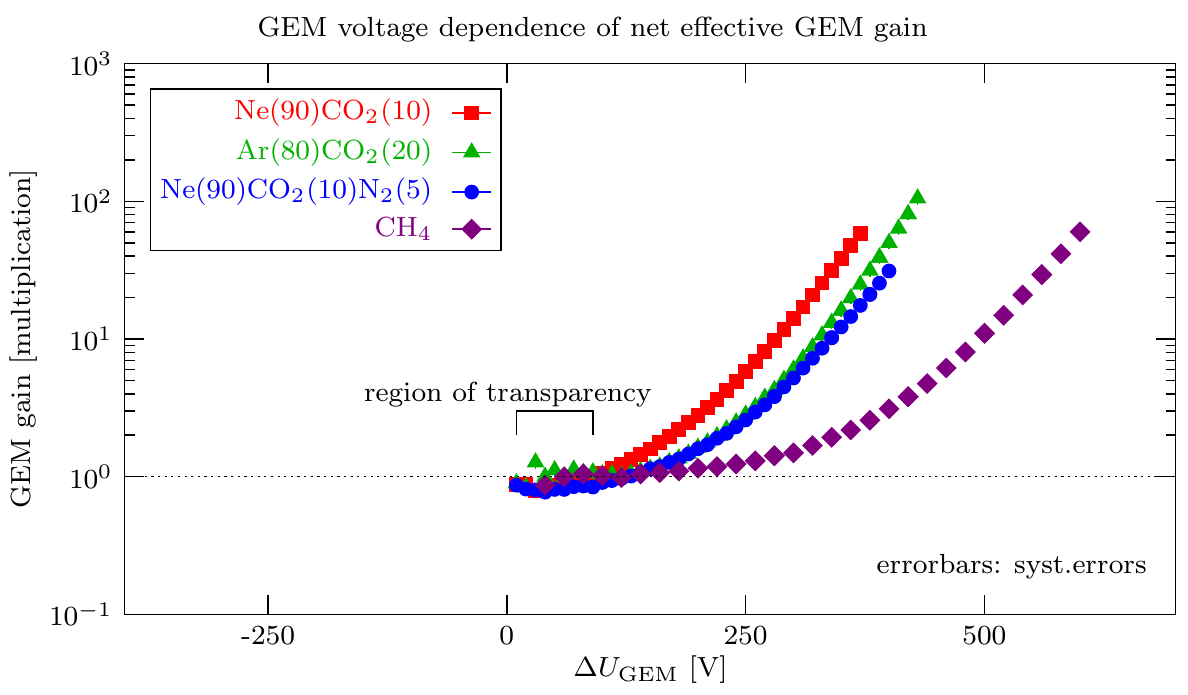}
\end{center}
\caption{(Color online) The determined net effective GEM gain curves as a function 
of GEM voltage $\Delta{U}_{\mathrm{GEM}}$, in various working gases.}
\label{gemgains}
\end{figure}

The calibration of the MWPC gain $\gamma$ was implemented in two steps. First, the 
Sense Wire voltage dependence of $\gamma$ was quantified. This was done 
by setting the amplifier GEM foil to a large gain, of the order of 
net effective multiplication $50$, and by setting a large PE yield, of the 
order of $1$ PE per pulse. Then, a Sense Wire voltage scan was performed. 
Because of Eq.(\ref{gain}), the shape of the Sense Wire voltage dependence 
of the MWPC gain could be determined since $\nu_{\eff}$ and $\mu_{\p}$ was kept constant. 
Given the shape of the Sense Wire voltage dependence of the MWPC gain 
$\gamma$, its absolute normalization was then obtained in a setting with 
transparent GEM (effective gain $\approx 1$), large PE yield ($\approx 1$), 
and a large Sense Wire voltage setting using the identity
\begin{eqnarray}
\frac{\sigma_{h}^{2}-\sigma_{g}^{2}}{\mu_{h}-\mu_{g}} & = & \frac{\sigma_{\f}^{2}+\mu_{\f}^{2}}{\mu_{\f}}
\label{gaindet}
\end{eqnarray}
which follows from Eq.(\ref{mom}), where in the present situation one has 
$\f=e_{\gamma}$ since the GEM did not amplify in such calibration runs. 
Note that the PE yield $\nu_{\eff}$ cancels in Eq.(\ref{gaindet}). 
Using now $\gamma=\mu_{e_{\gamma}}$ and $\sigma_{e_{\gamma}}/\mu_{e_{\gamma}}\approx 1$, the absolute MWPC 
gain can be estimated via the formula
\begin{eqnarray}
\gamma & \approx & \frac{1}{2}\,\frac{\sigma_{h}^{2}-\sigma_{g}^{2}}{\mu_{h}-\mu_{g}}
\label{mwpcgainnorm}
\end{eqnarray}
in such calibration setting. Note that due to the lower signal to noise ratio 
without the GEM amplification, such absolute calibration runs were only possible 
at larger PE yields and at larger MWPC gain settings in order to maintain 
a good signal to noise ratio. The compatibility of 
the Sense Wire voltage dependence of the absolute MWPC gain values were 
cross-checked with the relative calibration curve of the MWPC gain 
in a couple of extremal settings at large PE yield and large MWPC gain. 
The result of the MWPC gain calibration procedure is shown in Fig.~\ref{mwpcgains}. 
After the MWPC calibration procedure, the net effective GEM gain can always 
be calculated in a given setting via Eq.(\ref{gain}). The obtained 
net effective GEM gain curves are shown in Fig.~\ref{gemgains}. It is seen 
that in the region $\Delta{U}_{\mathrm{GEM}}\approx 10-90\,\mathrm{V}$ the 
net effective GEM gains are $\approx 1$ in all the working gases, i.e.\ the 
GEM foil becomes transparent to electrons and no multiplication takes place. 
In our experimental runs the transparent setting was thus defined as 
$\Delta{U}_{\mathrm{GEM}}=50\,\mathrm{V}$ on the electrodes.

\section{Results on GEM response distributions}
\label{measurements}

\begin{figure*}[!ht]
\begin{center}
\includegraphics{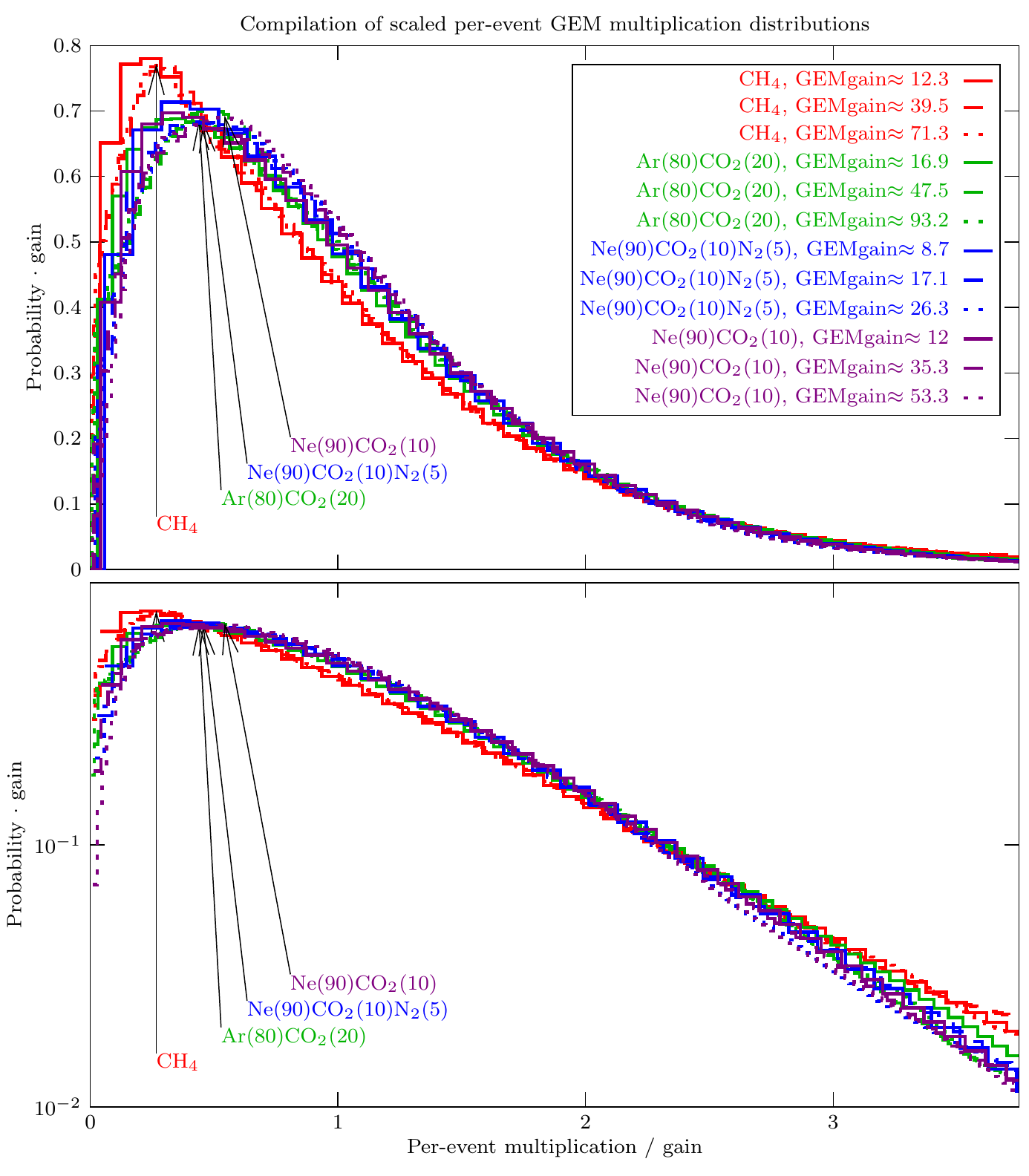}
\end{center}
\caption{(Color online) Top panel: the reconstructed shapes of the net effective 
$1$-electron multiplication distribution in GEM avalanche processes in 
various working gases at various net GEM effective gains. 
In a given working gas, very little gain dependence 
of the response shapes is seen. On the other hand, there is a systematic 
shape difference in different working gases. The closest to exponential 
shape is seen in CH${}_{4}$, whereas the most suppressed low amplitude 
region is observed in Ne(90)CO${}_{2}$(10) mixture. Bottom panel: the 
same distributions in logarithmic scale.}
\label{results1}
\end{figure*}

\begin{figure*}[!ht]
\begin{center}
\includegraphics{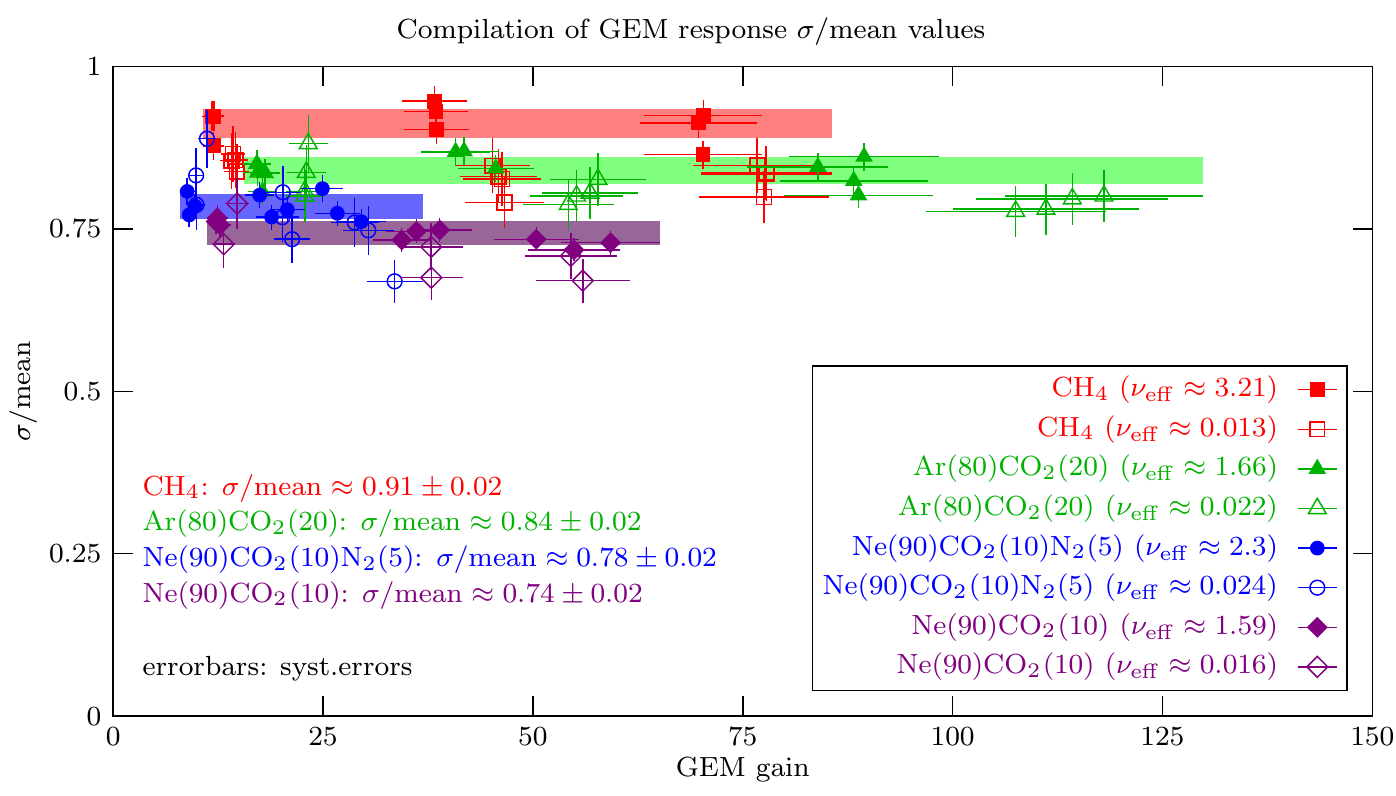}
\end{center}
\caption{(Color online) 
The reconstructed sigma-over-mean ratios for the $1$-electron multiplication distributions 
using the method of moments. In a given working gas, very little gain dependence 
of the sigma-over-mean ratio is seen. On the other hand, there is a systematic 
sigma-over-mean change between different working gases. The largest sigma-over-mean 
ratio was seen in CH${}_{4}$, whereas the smallest was observed in 
Ne(90)CO${}_{2}$(10) mixture. The measurement results with low PE yield ($\nu_{\eff}\approx 10^{-2}$) 
as well as with large PE yield ($\nu_{\eff}\approx 1$) are also shown for consistency check purposes. 
The measurements with the large PE yield are more reliable due to their much 
smaller systematic uncertainty in terms of PE yield determination. The errorbars 
as well as the thickness of the colored bands indicate systematic errors.}
\label{results2}
\end{figure*}

The measurement results of the $1$-PE response analysis is shown in Fig.~\ref{results1} and \ref{results2}. 
In Fig.~\ref{results1} the reconstructed net effective GEM multiplication 
distributions for single incoming electron, in different working gases and 
at various net effective gain settings are summarized. Fig.~\ref{results2} shows the sigma-over-mean 
ratios, obtained via the simpler method of moment reconstruction. Both 
results show that in a given working gas the shape of the GEM 
multiplication distribution has very little gain dependence in the studied 
region of net GEM effective gains, i.e.\ above about $15$. On the other hand, 
there is a non-negligible working gas dependence of the multiplication 
distributions. In pure CH${}_{4}$ the multiplication response is rather 
exponential-like, with a sigma-over-mean close to $1$. In the mixture 
Ar(80)CO${}_{2}$(20) a substantial deviation from the exponential distribution 
at low amplitudes is observed, with a decreased sigma-over-mean ratio. 
The mixture Ne(90)CO${}_{2}$(10)N${}_{2}$(5), to be used as 
working gas in the ALICE experiment \cite{alicetpc}, shows an even more 
suppressed yield of low amplitude GEM responses, resulting in a
smaller sigma-over-mean ratio. The mixture Ne(90)CO${}_{2}$(10) was seen 
to provide the best intrinsic response resolution, i.e.\ the smallest 
sigma-over-mean ratio. The results correspond well to the expectation that in gases where
the mean free path of electrons between collisions is longer, there will be stronger
departure from the limiting exponential distribution. If the mean
free path is very short, the electron mostly loses memory of the energy in the
last collision.

The estimated net GEM effective gains carry about $10\%$ systematic errors 
inherited from the MWPC gain curve normalization estimation via Eq.(\ref{mwpcgainnorm}). That originates 
from the uncertainty of the true value of the sigma-over-mean ratio of the MWPC 
response distribution, estimated to be about at most $10\%$ below the 
value $1$.

The raw data for the Fourier based Poisson compound decomposition of the $1$-PE 
multiplication response distributions 
were obtained at large effective PE yield $\nu_{\eff}$ of the order of 
$1$ PE per pulse for an increased signal to noise ratio. The principle 
of that reconstruction was shown in Figs.~\ref{numeasurement},\ref{Fmeasurement},\ref{fmeasurement} 
and the MWPC unfolding method was shown in Fig.~\ref{respfunc}. 
The systematic error of this analysis originates from the estimation accuracy of $\nu_{\eff}$, 
which is better than $5\%$ in systematics, and that propagates directly 
into the shape accuracy of the reconstructed multiplication distributions.

For the determination of the intrinsic response resolution, i.e.\ of 
the sigma-over-mean ratio of GEM multiplication distribution could be performed 
both at the low PE yield ($\nu_{\eff}\approx 10^{-2}$) and large PE yield 
($\nu_{\eff}\approx 1$) limit, using 
the formula Eq.(\ref{sigmapermeanpfinal}). The systematic errors of the estimated 
effective PE yield $\nu_{\eff}$ is smaller in the large $\nu_{\eff}$ limit, 
since it can be determined from the Fourier spectrum of the raw data in a 
relatively model independent way as was shown in Fig.~\ref{numeasurement}. 
In the small PE yield limit the model fitting method as shown in 
Fig.~\ref{GaussGammaPoissonFit} was used, for which an approximate model 
assumption is necessary for the shape of the $1$-PE multiplication distribution, 
bringing in some extra systematics due to the slight effect of the extrapolation uncertainty. 
Therefore, the measurement results using the large PE yield raw data is more 
accurate, having about $2.5\%$ systematic error, originating from the 
$5\%$ systematics of the $\nu_{\eff}$ determination.

\begin{figure*}[!ht]
\begin{center}
\includegraphics{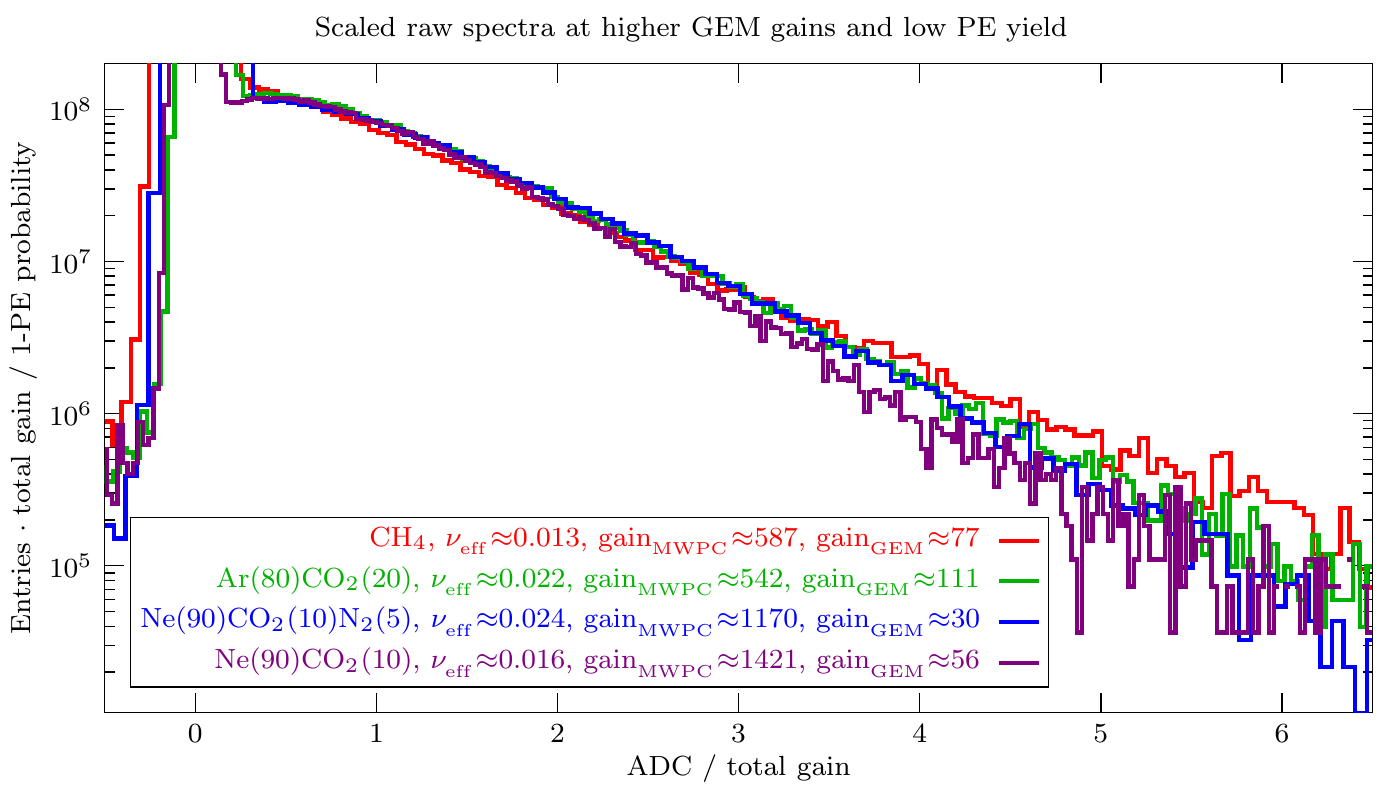}
\end{center}
\caption{(Color online) 
The raw amplitude distributions scaled appropriately with the gains and 
PE yields, overlayed on each-other. No sub-exponential tail is visible in the 
raw spectra, probably due to the moderate GEM gains.}
\label{results3}
\end{figure*}

Besides the behavior of the low amplitude region, the large amplitude tail 
of the GEM response distribution is also interesting. For instance, one can 
ask whether a large amplitude cutoff in the exponential-like tail is visible 
experimentally. The large amplitude behavior is quantified in Fig.~\ref{results3}: 
the raw amplitude spectra are shown there, scaled appropriately with the 
gains and PE yields, overlayed on each-other. No sub-exponential tail is 
seen in the spectra, possibly because of the applied not too large GEM gains. 
Note that for studying the large amplitude tails the 
raw spectra provide sufficient information as the effect of the MWPC becomes 
less and less important there, and therefore unfolding is not needed.

\section{Conclusions}
\label{conclusion}

In this paper a study on the single electron multiplication distribution 
in GEM foils has been presented. The experimental configuration based on 
photoelectron injection was outlined. The analysis methodologies for 
excluding detector effects such as multiple photoelectron contribution 
were detailed, and various cross-checks were shown. The 
multiplication distributions thus obtained show a deviation from exponential: low 
multiplication responses are suppressed in comparison to an 
exponential multiplication distribution of a low field limit avalanche process. 
This improves the intrinsic detector resolution of GEM based detectors, 
the sigma-over-mean ratio, in comparison to traditional MWPC detectors, if 
significant loss of the initial electron is not present in the system. 
The shapes of the multiplication distributions were seen to have negligible gain dependence 
throughout the effective net gain range of about $15$ to $100$. On the other hand, a dependence on 
the working gas was observed. A sigma-over-mean ratio down to $0.75$ was 
measured in neon-based working gases. That value is significantly lower than 
in case of traditional MWPC based multiplication processes, which are known to 
show a more exponential-like behavior.

\section*{Acknowledgements}

The authors thank Ferenc Sikl\'er and Eric D.\ Zimmerman for reading of the manuscript and for constructive critics. 
This work was supported by the Momentum (``Lend\"ulet'') 
Program of the Hungarian Academy of Sciences under the grant 
number LP2013-60, as well as the J\'anos Bolyai Research Scholarship 
of the Hungarian Academy of Sciences.


\begin{thebibliography}{00}

 \bibitem{curran1949} S.~C.~Curran, J.~Angus, A.~L.~Cockroft, \textit{Phyl.~Mag.} \textbf{40} (1949) 929.
 \bibitem{raether1964} H.~Raether, Electron avalanches and breakdown in gases (Butterworths, London, 1964).
 \bibitem{alkhazov1969} G.~D.~Alkhazov et al, \textit{Nucl.~Instr.~Meth.} \textbf{75} (1969) 161.
 \bibitem{alkhazov1970} G.~D.~Alkhazov et al, \textit{Nucl.~Instr.~Meth.} \textbf{89} (1970) 155.
 \bibitem{zerguerras2009} T.~Zerguerras et al, \textit{Nucl.~Instr.~Meth.} \textbf{A608} (2009) 397.
 \bibitem{zerguerras2015} T.~Zerguerras et al, \textit{Nucl.~Instr.~Meth.} \textbf{A772} (2015) 76.
 \bibitem{sauli2016} F.~Sauli, \textit{Nucl.~Instr.~Meth.} \textbf{A805} (2016) 2.
 \bibitem{alicetpc} B.~Abelev et al (the ALICE Collaboration), Technical Design Report for the upgrade of the ALICE Time Projection Chamber, \texttt{CERN-LHCC-2013-020} (2013).
 \bibitem{garfieldpp} Garfield++\newline \texttt{http://garfieldpp.web.cern.ch/garfieldpp}
 \bibitem{varga2016} D.~Varga, \textit{Advances in High Energy Physics} \textbf{2016} (2016) 8561743.
 \bibitem{leopard} G.~Hamar, D.~Varga, \textit{Nucl.~Instr.~Meth.} \textbf{A694} (2012) 16.
 \bibitem{ccc1} D.~Varga, G.~Hamar, G.~Kiss, \textit{Nucl.~Inst.~Meth.} \textbf{A648} (2011) 163.
 \bibitem{ccc2} D.~Varga, G.~Kiss, G.~Hamar, G.~Benc\'edi, \textit{Nucl.~Inst.~Meth.} \textbf{A698} (2013) 11.
 \bibitem{chargeup} B.~Azmoun et al, Nuclear Science Symposium Conference Record vol.\ 6 (IEEE, 2006) pp 3847-3851.
 \bibitem{feller1968} W.~Feller, \textit{An Introduction to Probability Theory and Its Applications} 3rd edn (New York: Wiley, 1968).
 \bibitem{haight1967} F.~A.~Haight, \textit{Handbook of the Poisson Distribution} (New York: Wiley, 1967).
 \bibitem{blum2008} W.~Blum, W.~Riegler, L.~Rolandi, \textit{Particle Detection with Drift Chambers} (Berlin: Springer-Verlag, 2008).
 \bibitem{arfken1985} G.~Arfken, Convolution theorem 15.5 \textit{Mathematical Methods for Physicists} 3rd edn (Orlando, FL: Academic, 1985) pp 810-14.
 \bibitem{bracewell1999} R.~N.~Bracewell, Convolution theorem \textit{The Fourier Transform and Its Applications} 3rd edn (New York: McGraw-Hill, 1999) pp 108-12.
 \bibitem{rudin1987} W.~Rudin, Theorem 9.6 \textit{Real and Complex Analysis} 3rd edn (New York: McGraw-Hill, 1987) pp 182.
 \bibitem{gsl} The GNU Scientific Library\newline \texttt{http://www.gnu.org/software/gsl}
 \bibitem{laszlo2016} A.~L\'aszl\'o, Convergence and error propagation results on a linear iterative unfolding method (\textit{to appear in SIAM JUQ, 2016}) [\texttt{arXiv:1404.2787}].
 \bibitem{laszlo2015} A.~L\'aszl\'o, The LibUnfold library\newline \texttt{http://www.rmki.kfki.hu/\~{}laszloa/downloads/}\newline \texttt{libunfold.tar.gz}
 \bibitem{laszlo2012} A.~L\'aszl\'o, \textit{J.~Phys.~Conf.~Ser.} \textbf{368} (2012) 012043.

\end{thebibliography}
\end{document}